\documentclass{aa}

\usepackage{allrunes}
\usepackage{graphicx}
\usepackage{txfonts}
\usepackage{wasysym}
\usepackage{array}
\usepackage[dvipsnames]{xcolor}
\usepackage[pdftitle={Transit least-squares survey. III. A $1.9\,R_\oplus$ transit candidate in the habitable zone of Kepler-160 and a non-transiting planet characterized by transit timing variations}, colorlinks = true, breaklinks = true, citecolor = blue, linkcolor = blue, urlcolor = blue, pdfauthor = {Ren\'{e} Heller}]{hyperref}
\usepackage{esvect}  % \vv{}

\usepackage{subcaption}

\newcolumntype{C}[1]{>{\centering\let\newline\\\arraybackslash\hspace{0pt}}m{#1}}

\defcitealias{HRH19}{Paper~I} 

\defcitealias{HHR19}{Paper~II}

\begin{document}

\title{Transit least-squares survey}
\subtitle{III. A $1.9\,R_\oplus$ transit candidate in the habitable zone of Kepler-160 and a nontransiting planet characterized by transit-timing variations}
\titlerunning{Transit least-squares survey -- III. Kepler-160}
\author{Ren\'{e} Heller\inst{1}
\and
Michael Hippke\inst{2,3}
\and
Jantje Freudenthal\inst{4}
\and
Kai Rodenbeck\inst{4}
\and
Natalie~M. Batalha\inst{5}
\and
Steve Bryson\inst{6}
          }

   \institute{Max Planck Institute for Solar System Research, Justus-von-Liebig-Weg 3, 37077 G\"ottingen, Germany\\
   \email{heller@mps.mpg.de}
   \and
   Sonneberg Observatory, Sternwartestra{\ss}e 32, 96515 Sonneberg, Germany\\
   \email{michael@hippke.org}
   \and
   Visiting Scholar, Breakthrough Listen Group, Berkeley SETI Research Center, Astronomy Department, UC Berkeley
   \and
   Institut f{\"u}r Astrophysik, Georg-August-Universit{\"a}t G{\"o}ttingen, Friedrich-Hund-Platz 1, 37077 G{\"o}ttingen, Germany\\
   \email{jfreude@astro.physik.uni-goettingen.de}
   \and
   Department of Astronomy \& Astrophysics, University of California, Santa Cruz, 1156 High St., Santa Cruz, CA 95064, USA
   \and
   NASA Ames Research Center, Moffett Field, CA 94901, USA
             }

   \date{Received 16 October 2019; Accepted 3 May 2020}

% \abstract{}{}{}{}{} 
% 5 {} token are mandatory

\abstract{The Sun-like star Kepler-160 (KOI-456) has been known to host two transiting planets, Kepler-160\,b and c, of which planet c shows substantial transit-timing variations (TTVs). We studied the transit photometry and the TTVs of this system in our search for a suspected third planet. We used the archival Kepler photometry of Kepler-160 to search for additional transiting planets using a combination of our \texttt{W\={o}tan} detrending algorithm and our transit least-squares (\texttt{TLS}) detection algorithm. We also used the {\tt Mercury} $N$-body gravity code to study the orbital dynamics of the system in trying to explain the observed TTVs of planet c. First, we recovered the known transit series of planets Kepler-160\,b and c. Then we found a new transiting candidate with a radius of 1.91$^{+0.17}_{-0.14}$ Earth radii ($R_\oplus$), an orbital period of 378.417$^{+0.028}_{-0.025}$\,d, and Earth-like insolation. The \texttt{vespa} software predicts that this signal has an astrophysical false-positive probability of ${\rm FPP}_3=1.8\,{\times}\,10^{-3}$ when the multiplicity of the system is taken into account. Kepler vetting diagnostics yield a multiple event statistic of ${\rm MES}=10.7$, which corresponds to an ${\sim}85\,\%$ reliability against false alarms due to instrumental artifacts such as rolling bands. We are also able to explain the observed TTVs of planet c with the presence of a previously unknown planet. The period and mass of this new planet, however, do not match the period and mass of the new transit candidate. Our Markov chain Monte Carlo simulations of the TTVs of Kepler-160\,c can be conclusively explained by a new nontransiting planet with a mass between about 1 and 100 Earth masses and an orbital period between about 7 and 50\,d. We conclude that Kepler-160 has at least three planets, one of which is the nontransiting planet Kepler-160\,d. The expected stellar radial velocity amplitude caused by this new planet ranges between about 1 and $20\,{\rm m\,s}^{-1}$. We also find the super-Earth-sized transiting planet candidate KOI-456.04 in the habitable zone of this system, which could be the fourth planet.}


\maketitle
%
%-------------------------------------------------------------------

\section{Introduction}
\label{sec:introduction}

The star Kepler-160 (KIC\,7269974, KOI-456), at a distance of $3141_{-54}^{+56}$ light years \citep{2016A&A...595A...1G,2018A&A...616A...1G,2018A&A...616A...8A}, was almost continuously observed during the Kepler primary mission from 2009 to 2013. It is a Sun-like star with a radius of $R_{\rm s}=1.118_{-0.045}^{+0.015}$ solar radii ($R_\odot$), an effective temperature of $T_{\rm eff}=5471_{-37}^{+115}$\,K, and a resulting Sun-like luminosity of $1.01_{-0.05}^{+0.05}$ times the solar value.\footnote{For any values cited from the Gaia DR2, lower and upper uncertainty limits refer to the 16th and 84th percentile of the probability density over the parameter, respectively. This corresponds to a range of about one standard deviation around the best-fit value to within 0.13\,\%.} The star has been identified as a host of two transiting-planet candidates in very close orbits \citep{2011ApJ...736...19B} that were later statistically validated \citep{2014ApJ...784...45R} as planets Kepler-160\,b and c. Kepler-160\,b has a radius of 1.7 Earth radii ($R_\oplus$) and is in a ${\sim}4.3$\,d orbit, while Kepler-160\,c, with a radius of about 3.1\,$R_\oplus$, orbits the star with a period of ${\sim}13.7$\,d. Kepler-160\,c has been shown to exhibit transit-timing variations (TTVs) with a period of about 879\,d \citep{2016ApJS..225....9H,2018ApJS..234....9O}, whereas Kepler-160\,b does not seem to exhibit TTVs.

As part of our transit least-squares survey \citep{2019A&A...625A..31H,2019A&A...627A..66H}, we analyzed the light curve of Kepler-160 to search for additional transiting planets that have previously been missed. We also examined the TTVs of Kepler-160\,c with orbital simulations of the Kepler-160 planetary system in combination with a Markov chain Monte Carlo (MCMC) fitting procedure to the Kepler transit photometry to search for additional planets. Here we report the detection of a new periodic transit feature in the light curve that {might originate from} a $1.9\,R_\oplus$-sized planet receiving virtually Earth-like insolation from its Sun-like host star. We also show that the TTVs of Kepler-160\,c are neither caused by this new transit candidate nor by the previously known planet Kepler-160\,b and that there must be yet another nontransiting planet in this system.

In Sect.~\ref{sec:methods} we explain the methods we used for the light-curve detrending and transit search, for the MCMC-based system characterization, for the automated false-positive vetting, and for the TTVs study through $N$-body simulations. In Sect.~\ref{sec:results} we present our results, Sect.~\ref{sec:discussion} is dedicated to a discussion of our findings, and in Sect.~\ref{sec:conclusions} we summarize and conclude.

%**********************************************
%Fig. 1
\begin{figure*}
\centering
\includegraphics[angle= 0,width=0.7\linewidth]{KIC7269974_.pdf}
\caption{\textbf{(a)} Kepler CCD photometry of Kepler-160 from the Data Validation Report. The dotted line outlines the aperture mask used to extract the light curve. The cross denotes the target position, and the stars (with annotated KIC numbers) indicate the positions of stars that might act as contaminants of the target light curve. The color scale illustrates the difference between mean flux out-of-transit and in-transit of Kepler-160\,b (KOI-456.01) after normalizing by the uncertainty in the difference for each pixel. North and east are indicated with yellow lines. The Kepler CCD pixels are squares and appear stretched in this figure from the Data Validation Report. \textbf{(b)} The Pan-STARRS image shows Kepler-160 in the center. The image is properly oriented so that north coincides with the Kepler image in \textbf{(a)}. The distance of $6.3\,\arcsec$ {to one nearby star} is indicated with an arrow.}
\label{fig:pixelmap}
\end{figure*}
%**********************************************

\section{Methods}
\label{sec:methods}

\subsection{Light-curve detrending and transit search}
\label{sec:detrending}

We used the publicly available pre-search data conditioning simple aperture photometry (PDCSAP) flux of the star Kepler-160 of the Kepler Data Release 25 (DR25) for Kepler Quarters 1 - 17 \citep{2016ksci.rept....3T,2016AJ....152..158T}
%RA~=~19\,h 11\,min 05.65\,s, DEC~=~+42\,deg 52\,min 09.47\,s)
that is accessible through the Mikulski Archive for Space Telescopes (MAST)\footnote{\href{https://archive.stsci.edu}{https://archive.stsci.edu}}. This light curve contains about 60,000 brightness measurements of the star between 13 May 2009 and 11 May 2013, mostly with a cadence of 30\,min. The time frame of the data is the Barycentric Kepler Julian Day (BKJD), which is equal to the Barycentric Julian Date (BJD) subtracted by 2,454,833.0\,d. BKJD = 0 corresponds to January 1, 2009, 12:00:00 UTC.

In our search for new transiting planets around Kepler-160, we first masked any data from the light curve that were affected by transits of the previously known planets Kepler-160\,b and c. {We removed any variability on timescales longer than the expected transit duration using the \texttt{W\={o}tan} software package\footnote{\href{https://github.com/hippke/wotan}{https://github.com/hippke/wotan}, version 1.2} \citep{Hippke_2019}.} \texttt{W\={o}tan} offers more than a dozen detrending filters, and we decided to use Tukey's biweight filter \citep{mosteller1977data}, which  has been shown to yield the highest recovery rates of injected transits from simulated data \citep{Hippke_2019}. We chose to use a running window of three times the {duration of a central transit on a circular orbit} at any given trial orbital period to detrend the light curve from long-term stellar variability and from any systematic trends on timescales longer than this value. This width has been shown to yield the best compromise between the trend removal and a high recovery rate of synthetic transits from simulated noisy light curves \citep{Hippke_2019}.

Then we passed the resulting detrended light curve to the \texttt{TLS} algorithm\footnote{\href{https://github.com/hippke/tls/}{https://github.com/hippke/tls}, version 1.0.23} \citep{2019A&A...623A..39H}. We used an oversampling factor of five to increase our sensitivity at the cost of computational effort, and we used estimates of the stellar limb-darkening coefficients, mass, and radius from the revised Kepler Input Catalog \citep{2011AJ....142..112B,2017ApJS..229...30M} to adapt the \texttt{TLS} transit search function to Kepler-160. The key detection criteria for our search were the signal-to-noise ratio (S/N) and the signal detection efficiency of \texttt{TLS} \citep[SDE$_{\rm TLS}$;][]{2019A&A...623A..39H}. The S/N was calculated as $(\delta/\sigma_{\rm o})\,n^{1/2}$ with $\delta$ as the mean depth of the trial transit, $\sigma_{\rm o}$ as the standard deviation of the out-of-transit points, and $n$ as the number of in-transit data points \citep{2006MNRAS.373..231P}. For a detection to be significant, we chose a minimum S/N of 7.1, which has been shown\footnote{{This threshold was based on the assumption of white noise only, although Kepler data have many time-dependent noise components (red noise).}} to yield one false positive for a threshold-crossing event over the entire Kepler mission \citep{2002ApJ...564..495J}. For the SDE$_{\rm TLS}$ value we applied a minimum threshold of 9, which results in a false-positive rate $<10^{-4}$ in the limiting case of white noise \citep{2019A&A...623A..39H}.

We also inspected the light curve for evidence of an eclipsing binary. {First, our automatic search with \texttt{TLS} only accepted candidate transit sequences for which} the average depths of the odd and even transits agreed within $3\,\sigma$. {In the case of the candidate transit signal that we report here, we had two odd transits and a single even transit.} {Second}, we ensured that there was no evidence of a secondary eclipse at the $>3\,\sigma$ level compared to the local noise at half an orbital phase after the candidate transit. {Third, to relax the circular-orbit assumption,} we performed a TLS search for secondary eclipses with orbital phases ranging between 0.1 and 0.9 (the transit being at 0). Although any secondary eclipse should have the same period ($P$) as the transit signal that we found, we allowed for a small variation of $P$ in our TLS search. As a result, the strongest signal always had both SDE$_{\rm TLS}~<~3$ and S/N$~<~3,$ and so we did not find any signal of a secondary eclipse with statistical significance.

We also verified that the period of the transit signal differs sufficiently from any periodic modulation of the light curve due to stellar and possibly systematic variability. As a tool for inspection, we chose the Lomb-Scargle power spectrum \citep{1976Ap&SS..39..447L,1982ApJ...263..835S} algorithm for unevenly spaced data \citep{1989ApJ...338..277P} as implemented in the \texttt{astropy} software package\footnote{\href{https://docs.astropy.org/en/stable/api/astropy.timeseries.LombScargle.html}{https://docs.astropy.org/en/stable/api/astropy.timeseries.Lomb\\ Scargle.html}}.

{All of these automated search metrics, together with the Kepler PDCSAP flux, the \texttt{W\={o}tan} detrending curve, the phase-folded light curve, secondary eclipse search, etc. were displayed on a vetting sheet \citep[for an example, see][]{2019A&A...625A..31H}. Our vetting sheet for Kepler-160 is one of hundreds that we inspected in our TLS survey. Some of the aspects that typically prompted us to reject a candidate based on the vetting sheet after it had already passed the automated detection criteria stated above were the following: (1) the transit period was very close to a period of activity in the Lomb-Scargle power spectrum, (2) there was substantial periodic activity in the light curve, (3) there were long gaps in the data, (4) the transit depths were highly variable, (5) there were features in the phase-folded light curve with amplitudes comparable to the primary signal, and (6) the transit shape was not convincingly ``planet-like''. Many of these aspects are highly subjective, but we nevertheless considered them very helpful in rejecting a large part of the automated candidate detections. Kepler-160 passed all of these human-based interventions in addition to our automated search criteria.}

\subsection{MCMC transit fitting of the light curve}

Our post-detection characterization of this putative three-planet transiting system involved the MCMC sampler \texttt{emcee} \citep{2013PASP..125..306F}. As inputs, we provided the mid-times of the first transits and $P$ for all three planets. The free parameters of our three-planet transit model were $P$, the transit epoch ($T_0$), the planet-to-star radius ratio ($R_{\rm p}/R_{\rm s}$), and the transit impact parameter ($b$) for each planet as well as the stellar density ($\rho_{\rm s}$) and two limb-darkening coefficients for a quadratic limb-darkening law \citep{2013MNRAS.435.2152K} that we used as global parameters for all transits. Our MCMC analysis was initialized with 100 walkers, and each walker performed $300,000$ steps. The first half of each walk was discarded to ensure that we preserved only burned-in MCMC chains. {We visually checked that the Markov chains were sufficiently burnt-in.} We selected the burn-in time by inspecting the chains for trends and changes in the spread, and we selected the acceptance fraction using a trace plot of the chains \citep{Roy2020}.

We computed the planetary radii from the posterior distribution of $R_{\rm p}/R_{\rm s}$ and from the estimate of $R_{\rm s}$ from the Gaia DR2 \citep{2016A&A...595A...1G,2018A&A...616A...1G}, which was derived from a combination of Gaia photometry and parallax measurements \citep{2018A&A...616A...8A}. For the sake of a self-consistent data set, we used $T_{\rm eff}=5471_{-37}^{+115}$\,K from the Gaia DR2 to estimate the stellar luminosity (see Sect.~\ref{sec:alternative} for a discussion of $T_{\rm eff}$). From the posterior distributions of $\rho_{\rm s}$ and the orbital parameters and using the Gaia DR2 values for $R_{\rm s}$ and $T_{\rm eff}$ as priors, we calculated the top-of-the atmosphere stellar flux received by the new transiting planet candidate.

\subsection{False-positive vetting}

We inspected the Kepler CCD pixel maps from the Data Validation (DV) Report of Kepler-160 (available at the NASA Exoplanet Archive) to search for possible stellar contaminants of the light curve. Stars that are physically unrelated but apparently close to the target on the plane of the sky can contribute to the total flux received within the aperture selected on the Kepler CCDs. If these stars are, for example, members of an eclipsing background binary or triple system, or if the target star is accompanied by a grazing eclipsing stellar binary, then the resulting flux measurement in the Kepler photometry can lead to a false-positive detection of a planet-like transit signal.

Figure~\ref{fig:pixelmap}(a) shows the Kepler CCD image of Kepler-160 from Kepler Quarter 5 (Q5), during which the first transit of our new exoplanet candidate was observed. The color scale illustrates the S/N for the previously known planet Kepler-160\,b. The position of the target star is denoted with a cross and additional stellar sources are marked with stars. Two stars north of the target (i.e., to the lower right in that image) are within the aperture mask used for the extraction of the light curve, which is outlined by a dotted line. Three more stars south-east of the target (left on the image) are about 1.2 pixels away from the edge of the aperture mask {that was used to extract the light curve of Kepler-160.} Figure~\ref{fig:pixelmap}(b) shows a high-resolution image of the region around Kepler-160 from Pan-STARRS, which we used {as an additional tool to assess the possibility of} stellar contamination.

{We used the \texttt{lightkurve} software \citep{2018ascl.soft12013L}\footnote{\href{https://docs.lightkurve.org}{https://docs.lightkurve.org}} to create per-pixel S/N charts for the new transit signal from differential CCD images and to validate that the newly discovered transit signal comes from the Kepler-160 target pixel and not from any of these nearby sources (see Sect.~\ref{sec:contamination}). Briefly, the per-pixel S/N of the difference images is the difference between in-transit and out-of-transit flux in each pixel divided by the propagated uncertainty in each pixel \citep{2013PASP..125..889B}.}

{Because the edge length of a Kepler CCD pixel corresponds to $3.98\,\arcsec$ on the sky, we chose a value of $\rho~=~3.98\,\arcsec$ for the maximum aperture radius interior to which the signal must have been produced. This is a conservative estimate compared to the value of $0.5\,\arcsec$} used by \citet{2016ApJ...822...86M} to evaluate the false-positive probabilities (FPPs) of Kepler-160\,b and c. We used the \texttt{vespa} software \footnote{\href{https://github.com/timothydmorton/VESPA}{https://github.com/timothydmorton/VESPA}} \citep{2012ApJ...761....6M,2015ascl.soft03011M} to evaluate the FPP of the signal, for example, the probability that it was caused by a blend or stellar eclipsing binary or triple system. {For the \texttt{vespa} {\tt maxrad} value we adopted a value of $3.98\,\arcsec$, and we supplied \texttt{vespa} with a phase-folded light curve of our new transit candidate from which all the transits of the previously known planets Kepler-160\,b and c had been removed.}

In addition to the Pan-STARRS optical images of Kepler-160 and its celestial neighborhood, we also inspected high-resolution adaptive optics images\footnote{\href{http://roboaokepler.org/koi\_pages/KOI-456.html}{http://roboaokepler.org/koi\_pages/KOI-456.html}} of Kepler-160 obtained by the Robotic Laser-Adaptive-Optics imaging survey \citep{2014ApJ...791...35L}. These images show no evidence of contaminants with a magnitude difference $<3$ within about $4\,\arcsec$ around Kepler-160. Although we cannot use this information for our analysis, which requires the input of the maximum possible radius for contamination, this information further increases our confidence that the transit signature that we discovered is not a false positive caused by contamination from an eclipsing or grazing binary or an unrelated transiting planet in the background.

{To assess the possibility that our new candidate might be caused by instrumental effects, such as the rolling-band effect or a combination of rolling-band and statistical artifacts, we vetted it using the \texttt{Model-Shift} software\footnote{\href{https://github.com/JeffLCoughlin/Model-Shift}{https://github.com/JeffLCoughlin/Model-Shift}} \citep{2017ksci.rept...15C}.}

%**********************************************
%Fig. 2
\begin{figure}
\centering
\includegraphics[angle= 0,width=1\linewidth]{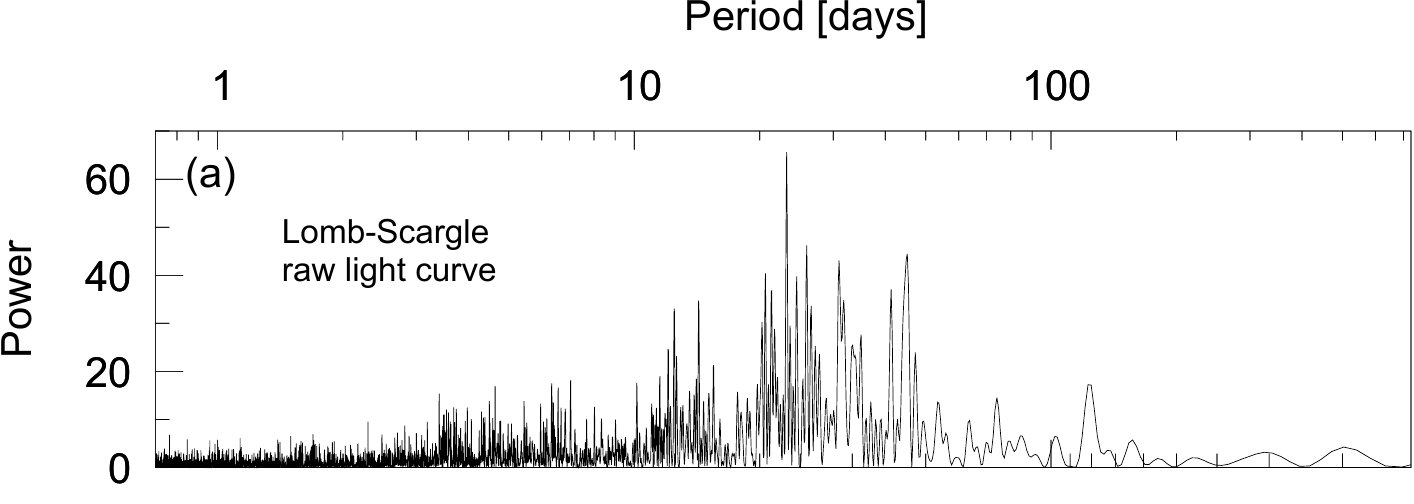}\\
\par
\vspace{-0.36cm}
\includegraphics[angle= 0,width=1\linewidth]{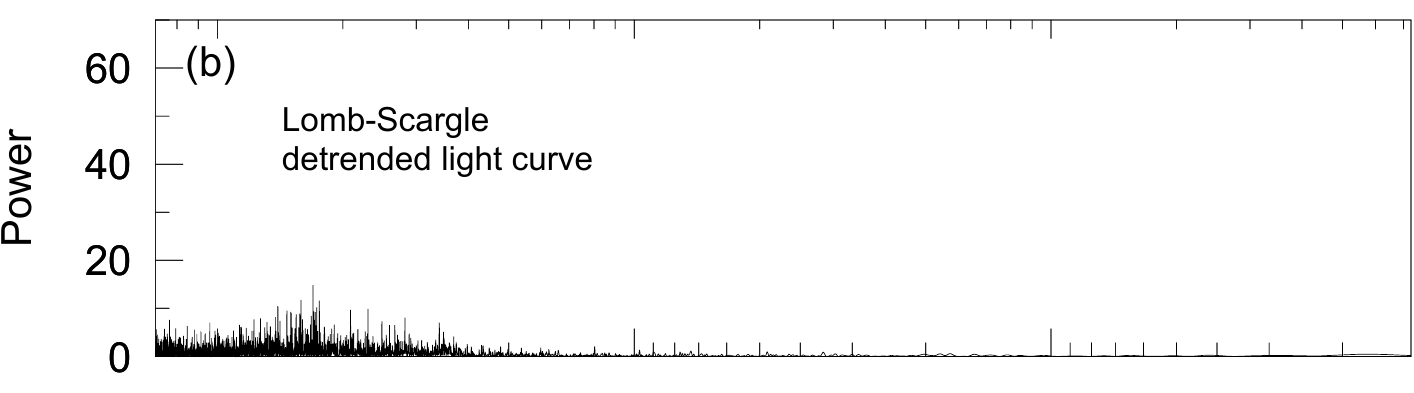}\\
\vspace{-0.2cm}
\includegraphics[angle= 0,width=1\linewidth]{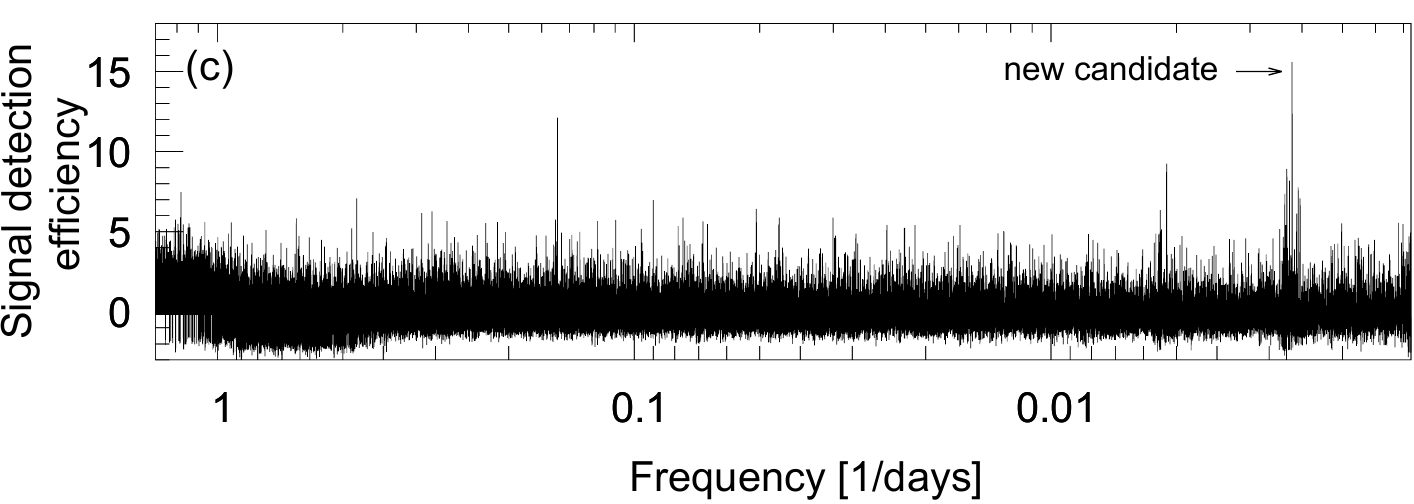}
\caption{Lomb-Scargle power spectrum and signal detection efficiency spectrum. \textbf{(a)} The Lomb-Scargle power spectrum of the raw PDCSAP Kepler light curve shows stellar (and possibly systematic) activity mostly at periods of between about 10\,d and 50\,d. The peak at roughly 22\,d might correspond to the stellar rotation period. \textbf{(b)} The activity has been removed from the light curve, which was detrended with a running biweight filter. \textbf{(c)} The signal detection efficiency spectrum produced with the \texttt{TLS} algorithm reveals a new candidate transit signal at a period of about 378\,d.}
\label{fig:periodogram}
\end{figure}
%**********************************************

\subsection{Transit-timing analyses of Kepler-160\,c}
\label{sec:TTVs}

We analyzed the TTVs of Kepler-160\,c with a dynamical model to determine the origin of the TTVs and to constrain the physical parameters of the perturber. Our model is similar to the photodynamical model applied by \citet{2018A&A...618A..41F,2019A&A...628A.108F}, except for the fact that it calculates the transit mid-times instead of the light curve. We used the $N$-body code {\tt Mercury} \citep{1999MNRAS.304..793C} to simulate the three-body (two planets plus one star) orbital dynamics of the Kepler-160 system over the duration of the observations. From the resulting orbital positions we calculated the transit times. The out-of-transit dynamics were numerically integrated with a typical time step of a twentieth of the period of the innermost planet and a time step of 0.05\,d near the transits. In a first attempt, we investigated whether Kepler-160\,b might be the cause of the observed TTVs of planet c. The results of this initial study were negative, and so we focused on additional planets in the system. 

% Table 1
\begin{table*}
\caption{Characterization of the transiting planets and the transit candidate in the Kepler-160 planetary system.}
\def\arraystretch{1.6}
\label{tab:system}
\centering
\begin{tabular}{lccc}
\hline
 Fitted model parameters & Kepler-160\,b & Kepler-160\,c & New transit candidate\\
\hline
Epoch, $T_0$ (BKJD) & 134.0320$^{+0.0028}_{-0.0023}$ & 171.4805$^{+0.0011}_{-0.0011}$ & 489.780$^{+0.031}_{-0.047}$\\
Period, $P$ (d) & 4.309397$^{+0.000013}_{-0.000012}$ & 13.699429$^{+0.000018}_{-0.000018}$ & 378.417$^{+0.028}_{-0.025}$\\
Planet-star radius ratio, $R_{\rm p}/R_{\rm s}$ & $0.01399_{-0.00032}^{+0.00044}$ & $0.03080_{-0.00053}^{+0.00091}$ & $0.0157_{-0.0011}^{+0.0014}$ \\
Impact parameter, $b$ & 0.19$^{+0.23}_{-0.13}$ & 0.26$^{+0.19}_{-0.15}$ & 0.34$^{+0.24}_{-0.22}$\\
S/N & 33.5 & 111.2 & 9.5\\
Number of transits (with data) & 451 (399) & 106 (95) & 3 (3)\\
SDE$_{\rm TLS}$ & 125 & 117.2 & 16.3\\
FPP$_{3}$ & {3.90~$\times$~10$^{-7}$} & {1.45~$\times$~10$^{-3}$} & {1.81~$\times$~10$^{-3}$} \\
\hline
 Derived parameters &  & &\\
 \hline
Planetary radius, $R_{\rm p}$ & 1.715$^{+0.061}_{-0.047}\,R_\oplus$ & 3.76$^{+0.23}_{-0.09}\,R_\oplus$ & 1.91$^{+0.17}_{-0.14}\,R_\oplus$ \\
Semimajor axis, $a$ & $10.62_{-0.71}^{+0.30}\,R_{\rm s}$ & $22.95_{-1.53}^{+0.66}\,R_{\rm s}$ & $209.7_{-13.9}^{+6.0}\,R_{\rm s}$ \\
 & $0.05511_{-0.0037}^{+0.0019}$\,AU & $0.1192_{-0.0080}^{+0.0040}$\,AU & $1.089_{-0.073}^{+0.037}$\,AU \\
\hline
\end{tabular}\\
\tablefoot{The values for $T_0$, $P$, $R_{\rm p}/R_{\rm s}$, and $b$ were obtained from an MCMC fitting procedure to the entire detrended Kepler light curve and using linear ephemerides. S/N values were computed from the detrended light curve after MCMC fitting and phase-folding. The number of transits and SDE$_{\rm TLS}$ were calculated with \texttt{TLS}. The FPP$_3$ values were computed with \texttt{vespa} and take the statistical multiplicity boost of this system into account. Planetary radii were derived from the fitted values for ($R_{\rm p}/R_{\rm s}$) and the stellar radius of $R_{\rm s}=1.118_{-0.045}^{+0.015}\,R_\odot$ from Gaia DR2. The $a/R_{\rm s}$ ratios were computed from the global fit of $\rho_{\rm s}=0.86_{-0.16}^{+0.08}\,\rho_\odot$ and the fitted values for $P$, respectively. The values of $a$ units of AU were calculated from $a/R_{\rm s}$ and using the Gaia DR2 value of $R_{\rm s}$. The stellar limb-darkening parameters were found to be $u_1=0.88_{-0.16}^{+0.08}$ and $u_2=0.39_{-0.08}^{+0.08}$.}
\end{table*}

%**********************************************
%Fig. 3
\begin{figure*}
\centering
\includegraphics[angle= 0,width=0.995\linewidth]{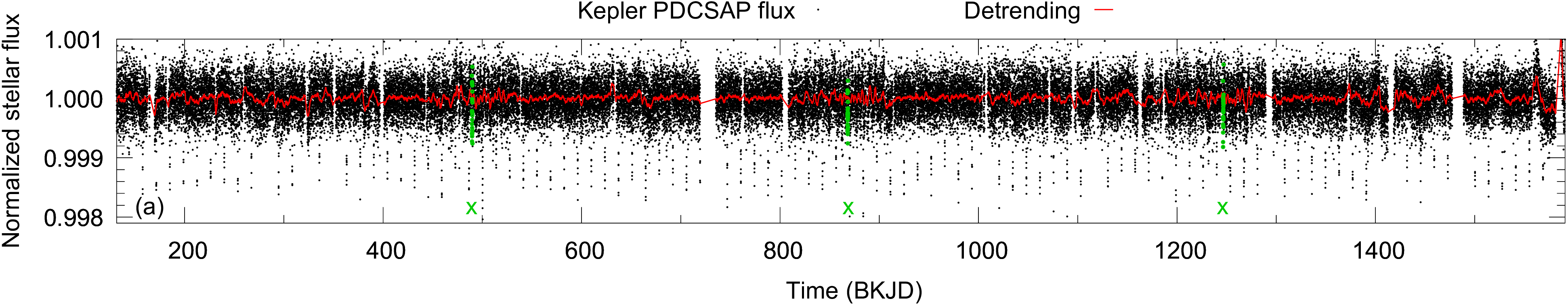}\\
\par
\vspace{0.3cm}
\includegraphics[angle= 0,width=0.37\linewidth]{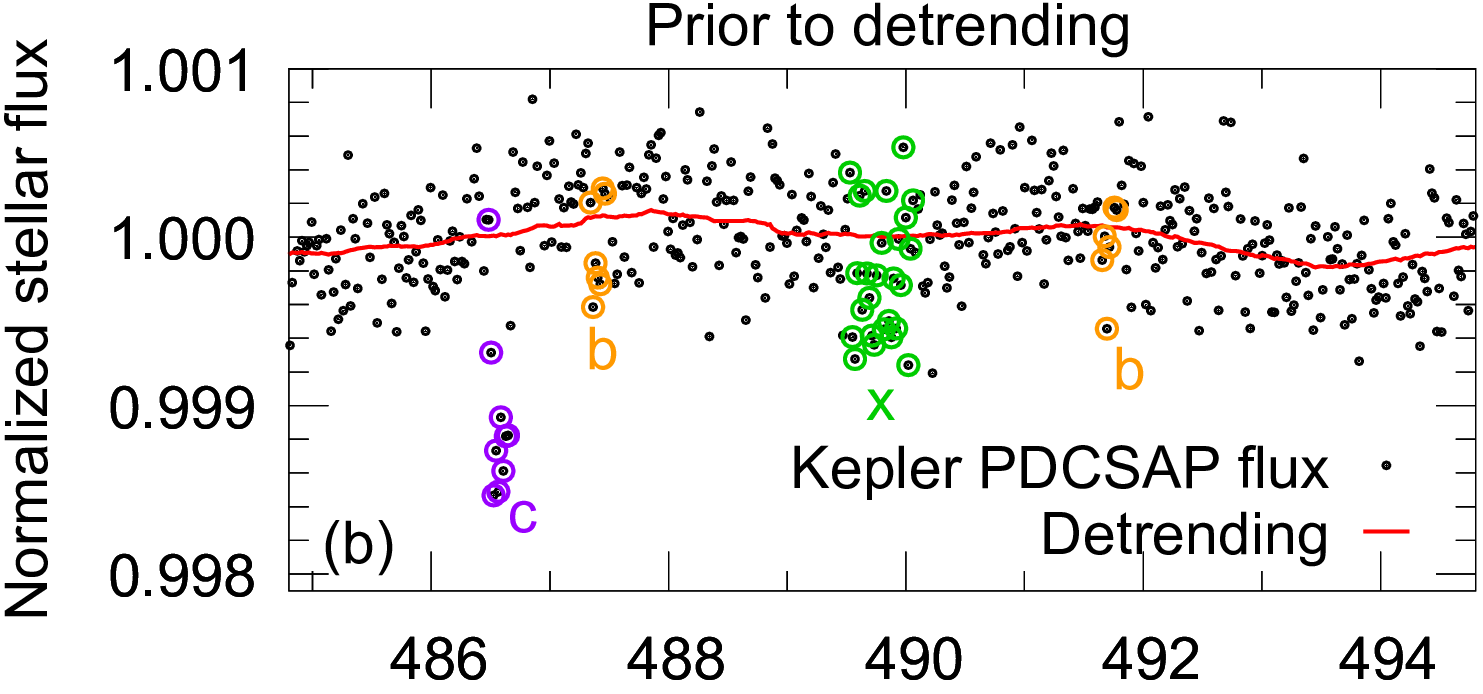}
\hspace{-0.2cm}
\includegraphics[angle= 0,width=0.30\linewidth]{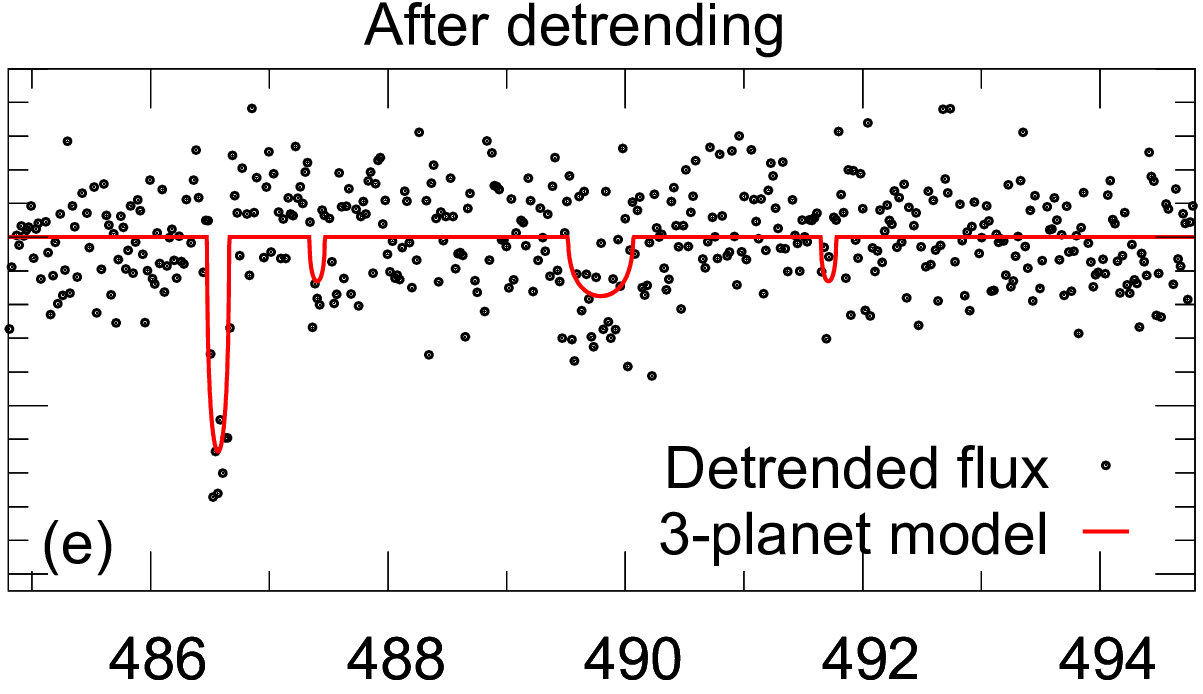}
\hspace{0.2cm}
\includegraphics[angle= 0,width=0.305\linewidth]{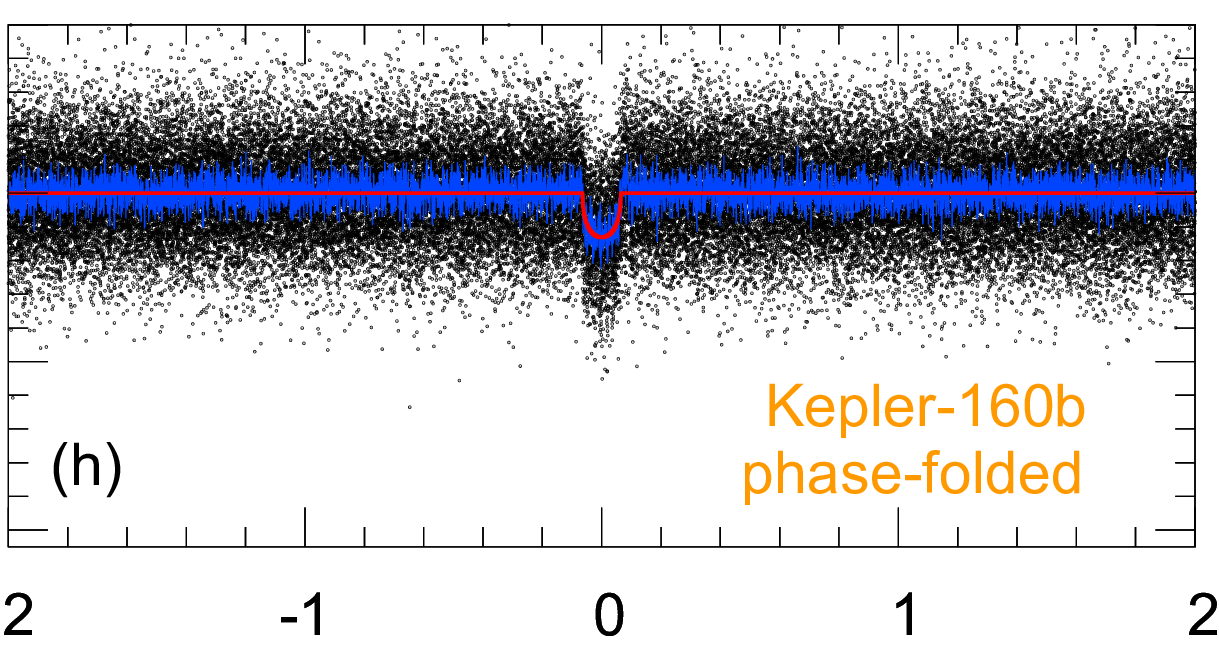}\\
\includegraphics[angle= 0,width=0.37\linewidth]{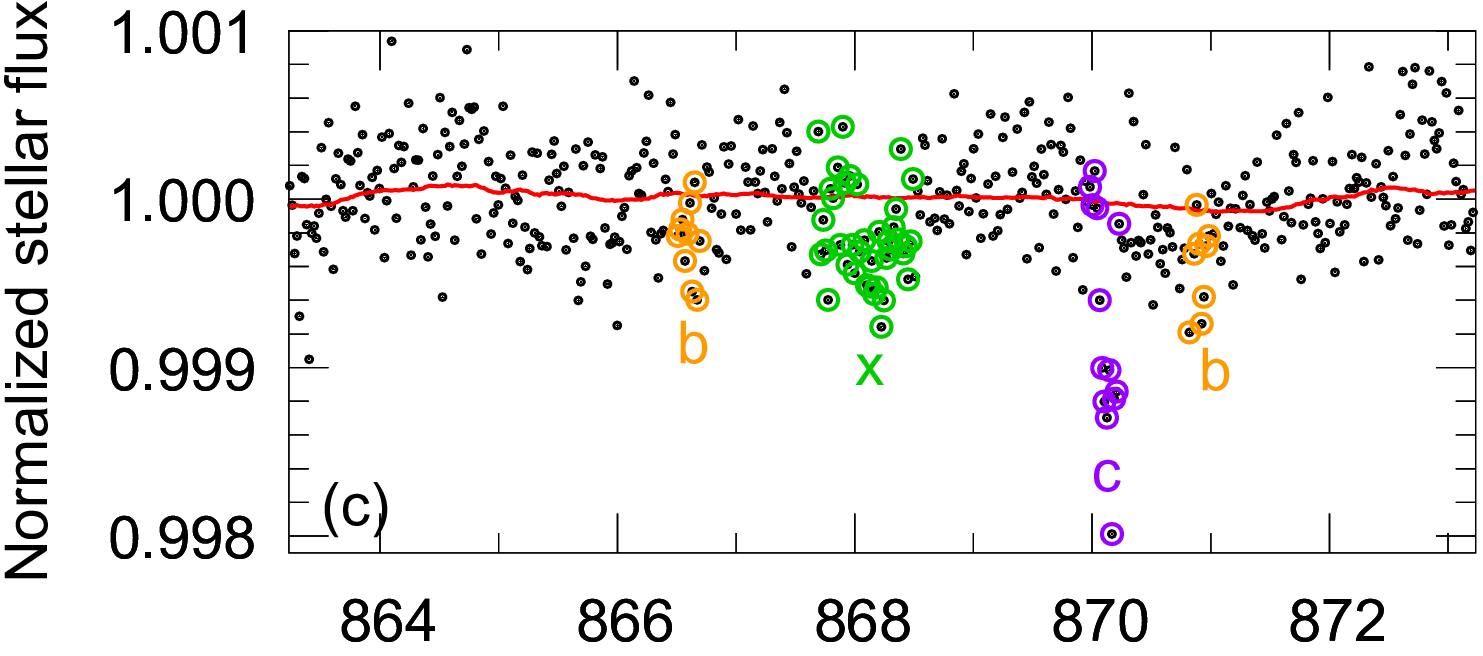}
\hspace{-0.2cm}
\includegraphics[angle= 0,width=0.30\linewidth]{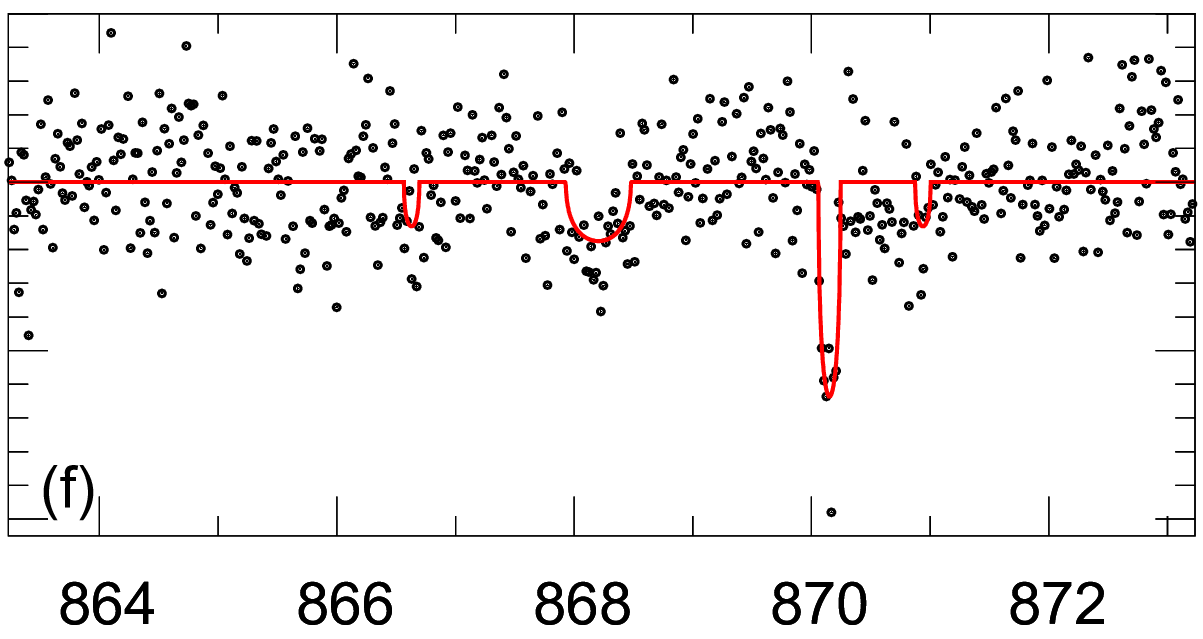}
\hspace{0.2cm}
\includegraphics[angle= 0,width=0.305\linewidth]{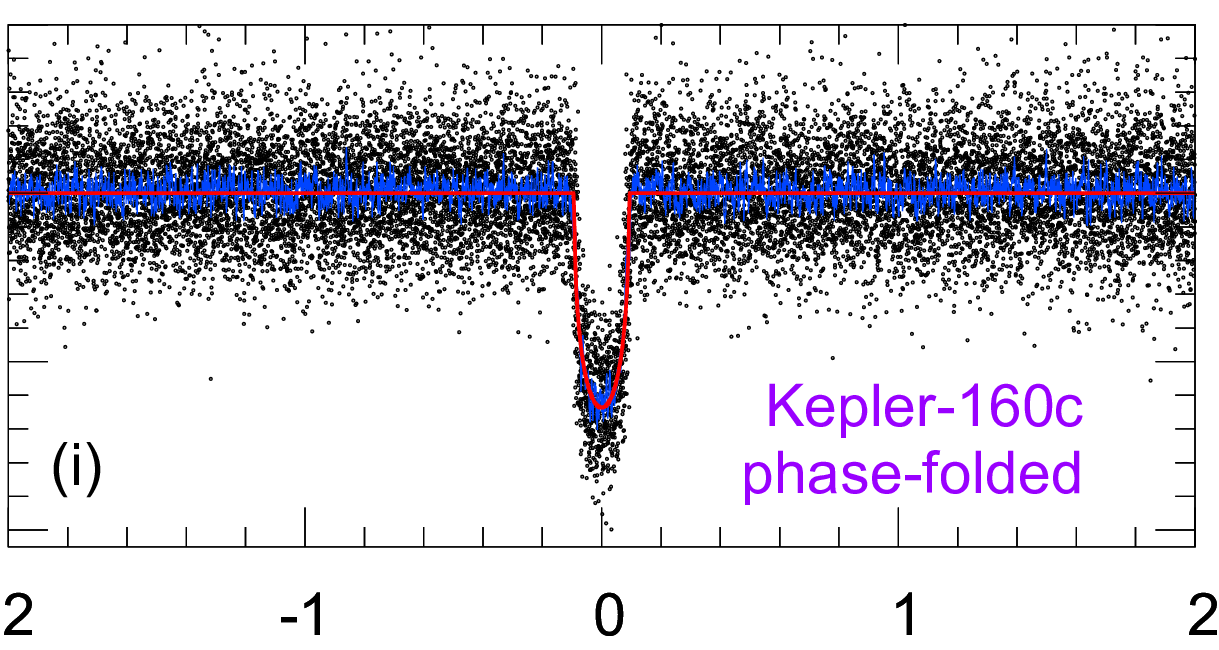}\\
\includegraphics[angle= 0,width=0.37\linewidth]{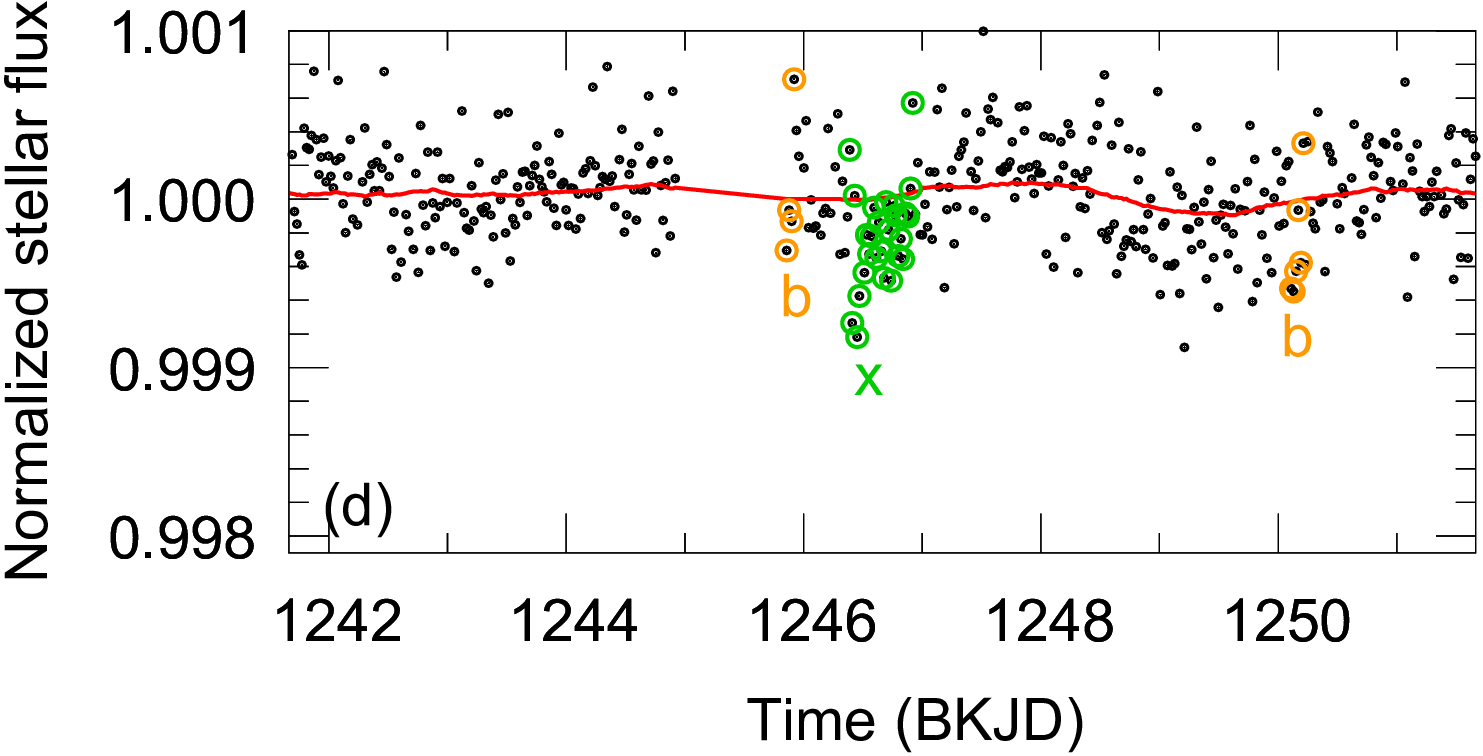}
\hspace{-0.2cm}
\includegraphics[angle= 0,width=0.30\linewidth]{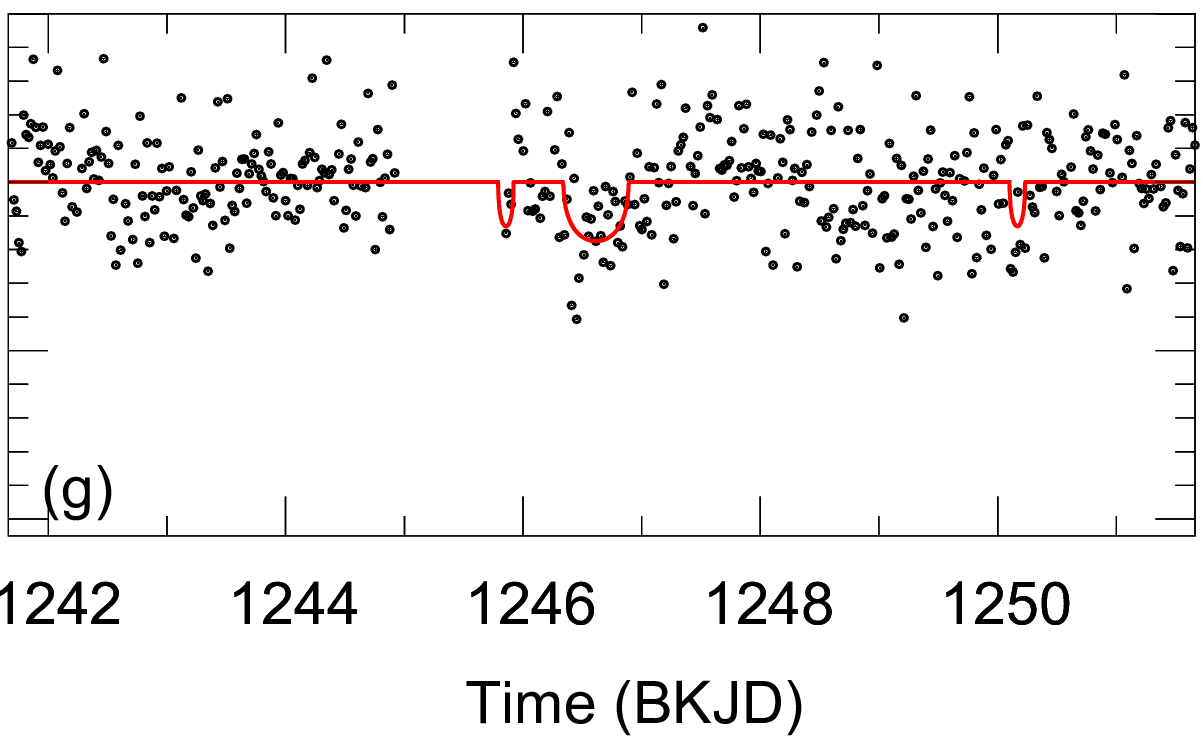}
\hspace{0.2cm}
\includegraphics[angle= 0,width=0.305\linewidth]{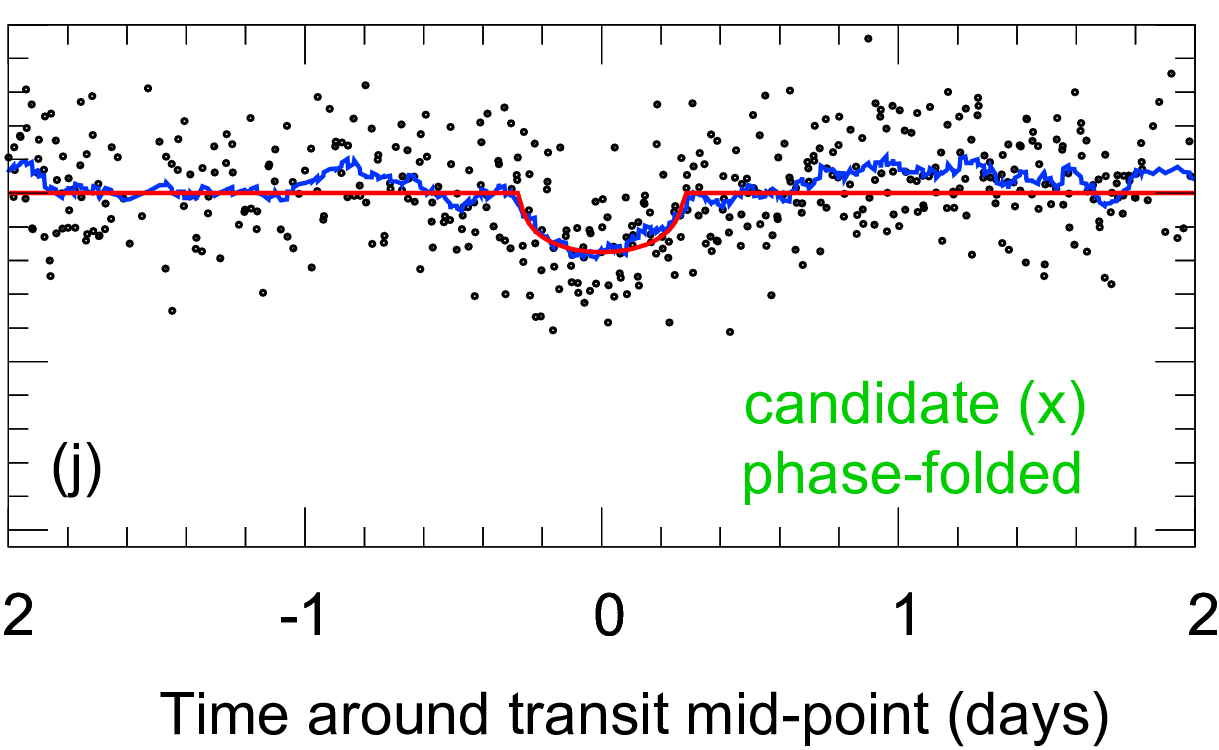}
\caption{Kepler photometry of Kepler-160. \textbf{(a)} Entire light curve of Kepler-160 with four years of data. Black dots illustrate the PDCSAP flux, and the red line shows our detrending as obtained with the biweight filter. All values have been normalized by the arithmetic mean of the respective Kepler quarter. Green data points highlight the in-transit flux measurements detected with \texttt{TLS} after detrending. Panels \textbf{(b)}-\textbf{(d)} show the three new candidate transits highlighted in green and labeled with an x. Transits of the previously known planets Kepler-160\,b and c are labeled in violet and orange and labeled b and c, respectively. Black dots refer to the Kepler PDCSAP flux without transits, and the red line shows the detrending function obtained with a running biweight window of 2.2\,d length. Panels \textbf{(e)}-\textbf{(g)} illustrate the detrended light curves and the best-fit model photometry of the system centered around the candidate transits. Panels \textbf{(h)}-\textbf{(j)} show the phase-folded light curves of Kepler-160\,b, Kepler-160\,c, and the new candidate, respectively. The red line shows the respective best-fit transit model and the 17-bin running mean is plotted as the blue line.}
\label{fig:lightcurve}
\end{figure*}
%**********************************************

Then we explored the parameter space of our model to identify the regions that maximize the likelihood of the set of parameters given the data. For this purpose, we again used \texttt{emcee} \citep{2013PASP..125..306F} to deduce the posterior distributions of the planetary parameters. Our focus was on the orbital period and mass of the unknown perturber of Kepler-160\,c. In order to generate the simulated TTVs of Kepler-160\,c from our {\tt Mercury} simulations, we used the following jump parameters for the MCMC fitting: the semimajor axis of planet c as well as the eccentricities, arguments of periastron, and the true anomalies of planets c and the test planet. The inclinations of both planets were fixed to $90^\circ$ (see Sect.~\ref{sec:coplanarity} for a discussion).

Because the TTV signal of Kepler-160\,c is periodic (see Sect.~\ref{sec:non}), we studied the TTVs caused by test planets near various orbital mean motion resonances (MMRs) with low integer ratios. All jump parameters had uniform priors with broad boundaries in order to avoid unphysical results. Starting values for the period of the perturber were calculated from the super-period given by the periodicity in the TTVs of Kepler-160\,c and its own period \citep[Eqs.~5 - 7 in][]{2012ApJ...761..122L}. The starting values of the remaining jump parameters were chosen manually so that the model approximately matched the data and were then optimized according to the MCMC procedure.

\section{Results}
\label{sec:results}

\subsection{Transit detection}

Figure~\ref{fig:periodogram} shows the results of our \texttt{TLS} search after masking out the known transits of Kepler-160\,b and c. Panel (a) illustrates the Lomb-Scargle power spectrum of the PDCSAP Kepler light curve prior to our own detrending, and panel (b) shows the spectrum of the detrended light curve. {A comparison between panels (a) and (b) shows the effect of the removal of stellar activity by \texttt{W\={o}tan} at frequencies ${\lesssim}~0.1\,{\rm d}^{-1}$, or equivalent at periods ${\gtrsim}~10$\,d.} Figure~\ref{fig:periodogram}(c) shows the SDE$_{\rm TLS}$ spectrum with a notable peak at a period of 378\,d. Any periodic activity in the Lomb-Scargle periodograms of panels (a) and (b) is separated by at least one order of magnitude in period from the 378\,d period of the suspicious transit signal. {As a consequence}, we do not find any clear signs of a connection between stellar activity and this candidate signal.

We also note a spurious peak at a period of 6.54206\,d, or a frequency of 0.15286\,d$^{-1}$ in the SDE$_{\rm TLS}$ spectrum of Fig.~\ref{fig:periodogram}(c). This value is not near any integer multiple of the period of our candidate signal. We have also verified that the spurious signal is not an alias of the new transit candidate signal (and vice versa). Moreover, the spurious signal remained in the SDE spectrum after the three transits of our new candidate were cut out from the light curve. We also split the Kepler light curve into two halves, which showed that the spurious signal is significantly more enhanced in the first half. We conclude that the spurious signal and the new transit candidate signal have independent origins. Instead, the spurious SDE peak at 0.15286\,d$^{-1}$ in Fig.~\ref{fig:periodogram}(c) coincides with a moderate peak in the LS power spectrum in Fig.~\ref{fig:periodogram}(a), which disappears after detrending in Fig.~\ref{fig:periodogram}(b). We interpret this as evidence that the 0.15286\,d$^{-1}$ signal is caused by stellar activity. We also inspected the phase-folded light curve of this spurious signal, and it looks only marginally consistent with a transit shape. Finally, its S/N of 5.6 (compared to 9.5 for our new candidate) also suggests that this signal might be caused by alternative noise sources rather than by a transit.

{Our detection of the candidate signal} with \texttt{TLS} gave us first estimates of $P$, $T_0$, the transit depth ($\delta$), and transit duration ($d$) of the candidate. The signal is related to a sequence of three periodic transit-like events distributed over Q5, Q9, and Q13. {Our initial measurements of the individual transits depths were obtained with \texttt{TLS} from the mean in-transit flux \citep{2019A&A...623A.137H,2019A&A...623A..39H}. The measured depths for the triple transit sequence are $207\,(\pm\,54)$\,ppm, $237\,(\pm\,46)$\,ppm, and $370\,(\pm\,41)$\,ppm, respectively. The mean value is $271$\,ppm, and the individual deviations are $-1.2\,\sigma$, $-0.8\,\sigma$, and $+2.4\,\sigma$, respectively.} We also searched for further periodic transit sequences beyond those of Kepler-160\,b, c, and the new candidate with \texttt{TLS} by masking the transits of the two planets and of the new candidate. This final search, however, did not yield any further planetary candidates with both S/N and SDE values higher than 6.

In Fig.~\ref{fig:lightcurve} we show (a) the entire Kepler photometry of Kepler-160, (b-d) close-up inspections of the three transits of the new candidate prior to detrending and (e-g) after detrending, and (h-j) the phase-folded transits of Kepler-160\,b and c and of the new candidate. The resulting system parameterization from \texttt{emcee} is shown in Table~\ref{tab:system}.

\subsubsection{Contamination from nearby sources}
\label{sec:contamination}

{As a first estimate for} the possible contamination of the light curve from the three sources visible in the southeast direction on the Kepler CCD pixel map (to the left of the target in Fig.~\ref{fig:pixelmap}), we explored the Gaia DR2 data and found that these three sources have Gaia magnitudes $m_{\rm G,con}~{\gtrsim}~20.5$ compared to $m_{\rm G,K160}=~14.6$ for Kepler-160. The Gaia bandpass is a good proxy for the Kepler bandpass. The magnitude modulus then gives the flux ratio ($F_{\rm K160}/F_{\rm con}$) between the sources as $m_{\rm G,con} - m_{\rm G,K160}~\approx~2.5\log_{10}(F_{\rm K160}/F_{\rm con})$. Equivalently, $F_{\rm K160}~=~F_{\rm con}~\times~10^{ (m_{\rm G,con}-m_{\rm G,K160})/2.5 }~=~229$. {In other words, the flux from the nearby stars is lower than one percent of that from Kepler-160, but it cannot be fully excluded that an eclipsing binary might mimic our transit signal with its depth of ${\sim}300$\,ppm.}

%**********************************************
%Fig. 4
\begin{figure}
    \begin{subfigure}[b]{.50\linewidth}
        \caption{Mean flux Q5}
        \vspace{-.2cm}
        \includegraphics[width=1\linewidth]{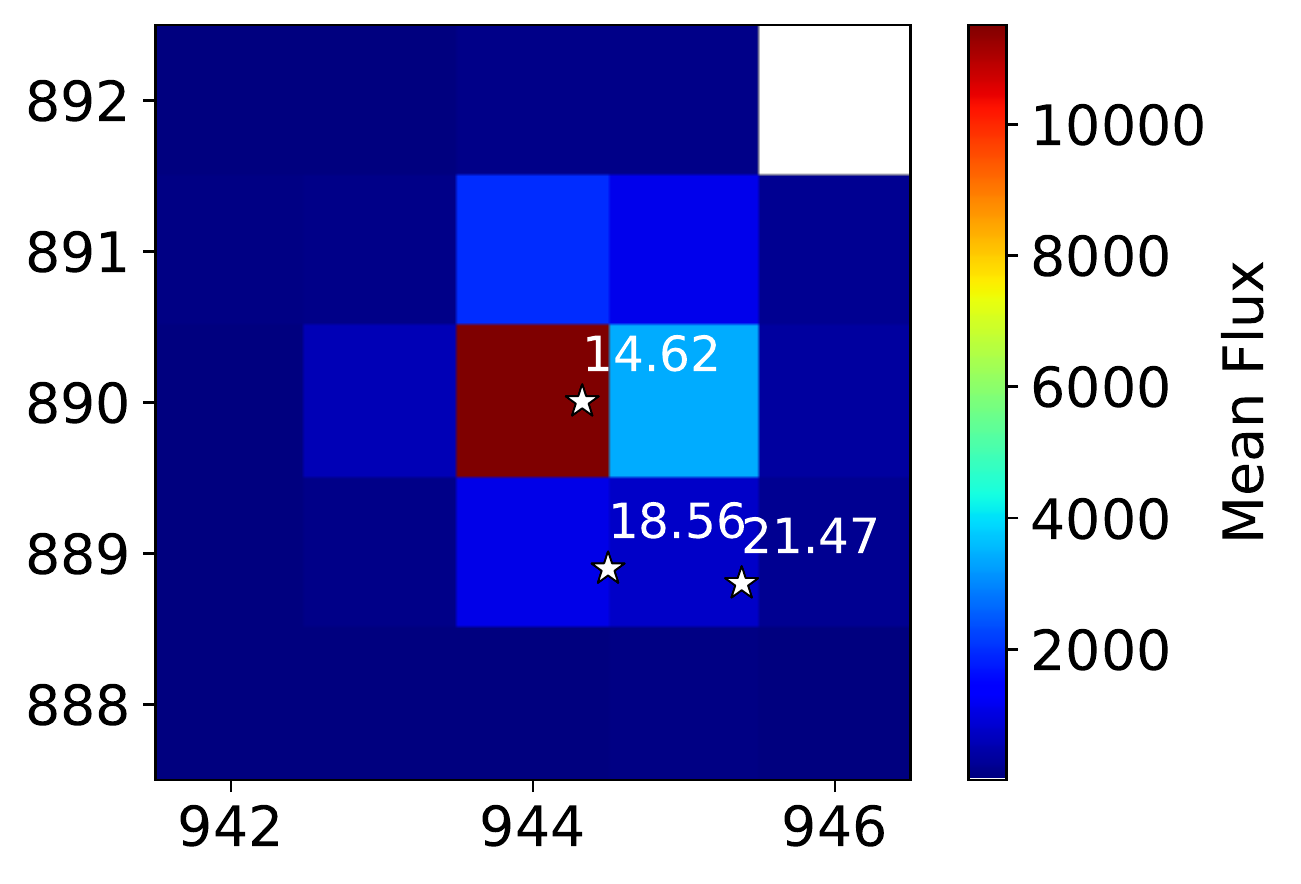}
        \label{fig:meanQ5}
    \end{subfigure}
    \hspace{0.15cm}
    \begin{subfigure}[b]{.43\linewidth}
        \caption{Difference S/N Q5}
        \vspace{-.2cm}
        \includegraphics[width=1\linewidth]{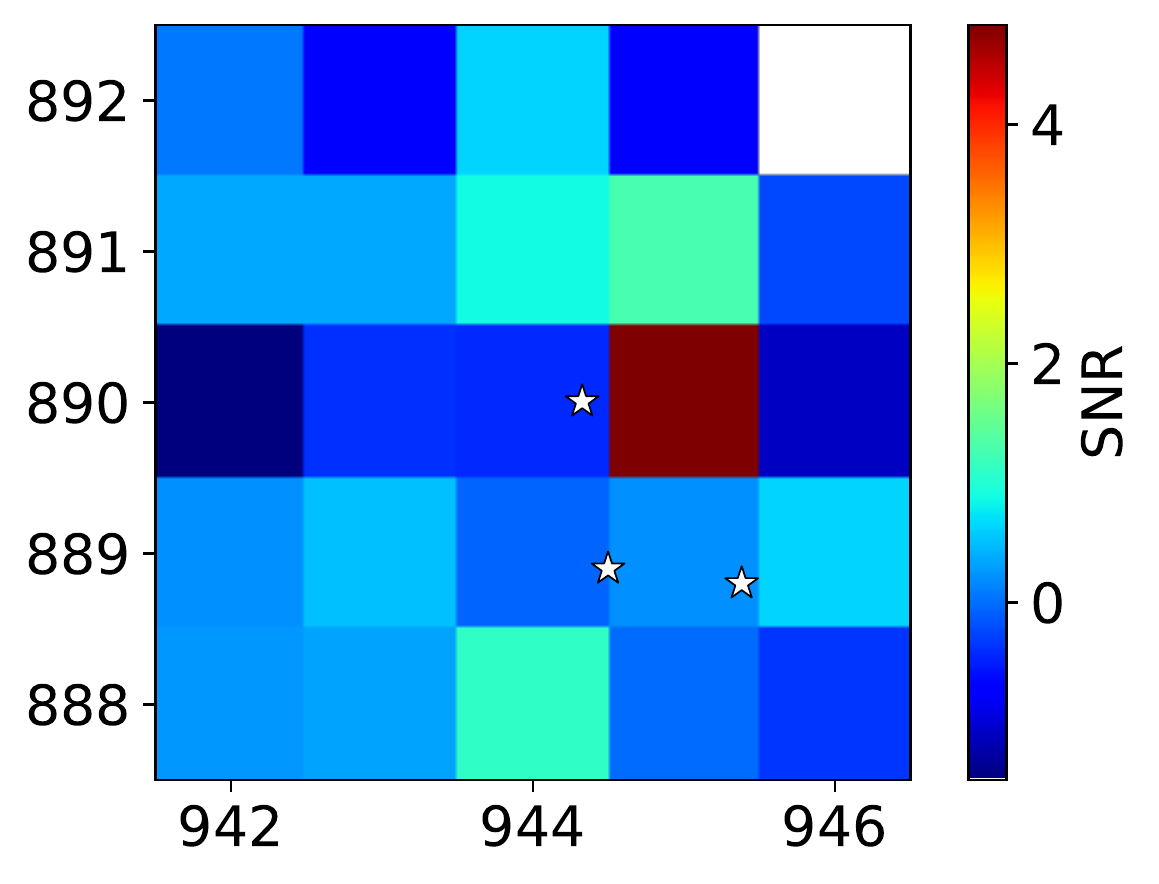}
        \label{fig:SNRQ5}
    \end{subfigure}
    
    \begin{subfigure}[b]{.495\linewidth}
        \caption{Mean flux Q9}
        \vspace{-.2cm}
        \includegraphics[width=1\linewidth]{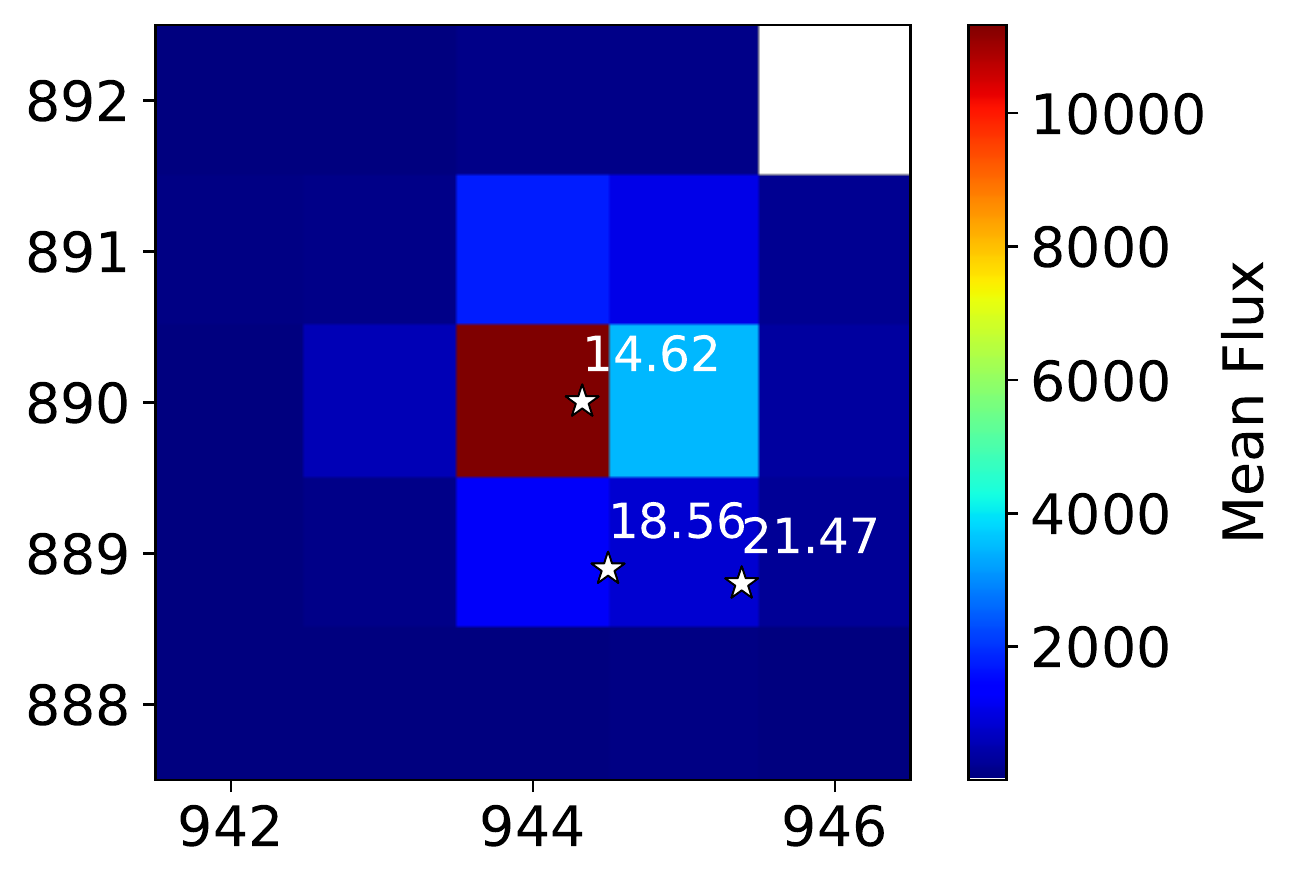}
        \label{fig:meanQ9}
    \end{subfigure}
    \hspace{0.15cm}
    \begin{subfigure}[b]{.465\linewidth}
        \caption{Difference S/N Q9}
        \vspace{-.18cm}
        \includegraphics[width=1\linewidth]{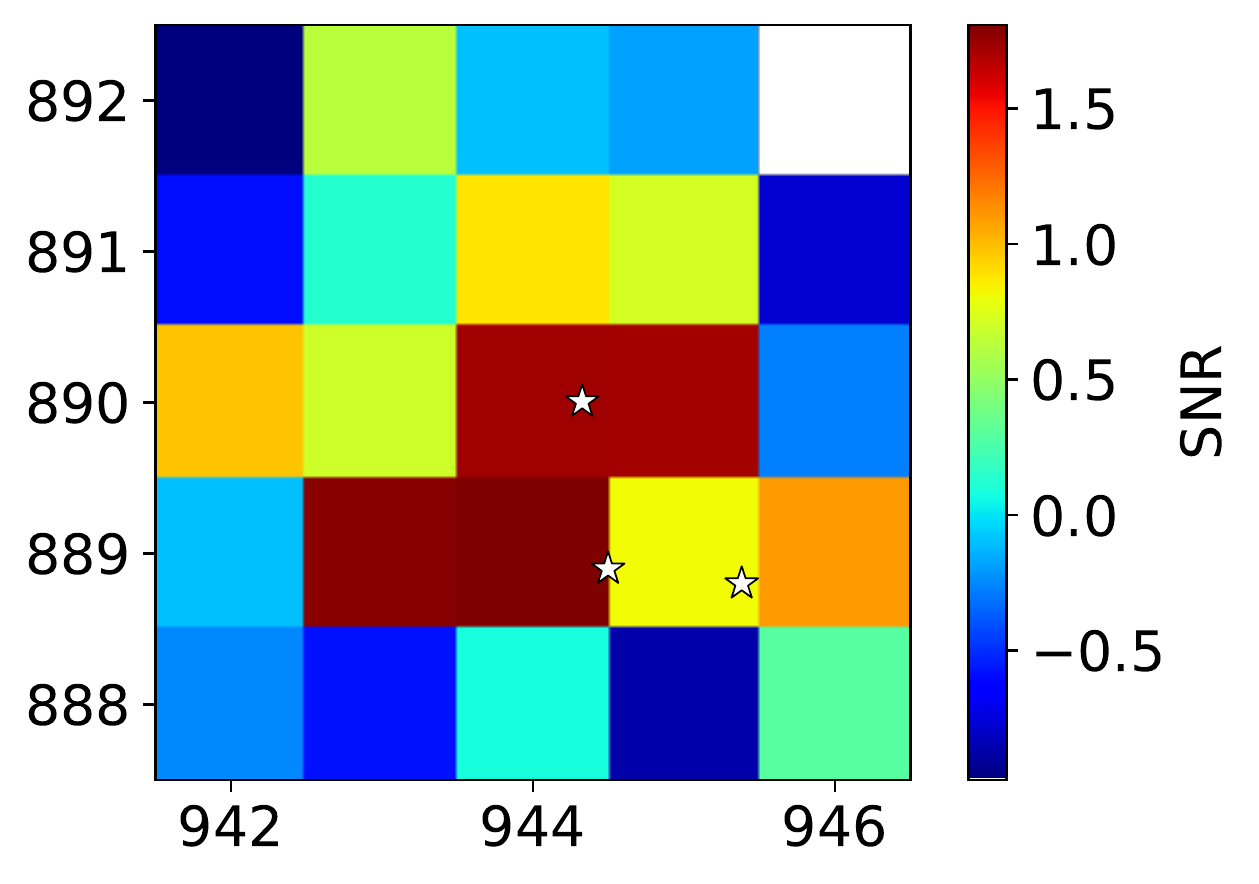}
        \label{fig:SNRQ9}
    \end{subfigure}

    \begin{subfigure}[b]{.495\linewidth}
        \caption{Mean flux Q13}
        \vspace{-.2cm}
        \includegraphics[width=1\linewidth]{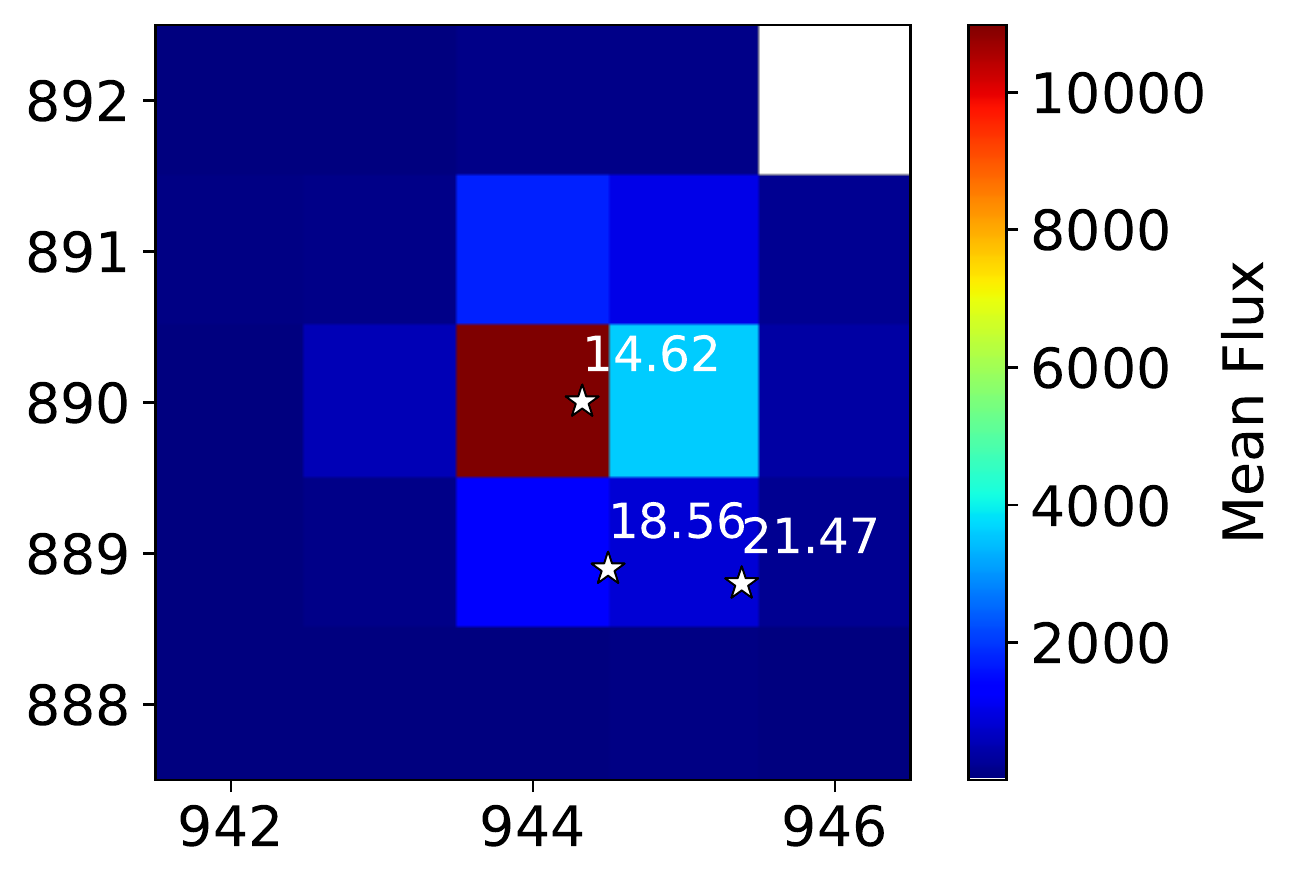}
        \label{fig:meanQ13}
    \end{subfigure}
    \hspace{0.15cm}
    \begin{subfigure}[b]{.470\linewidth}
        \caption{Difference S/N Q13}
        \vspace{-.18cm}
        \includegraphics[width=1\linewidth]{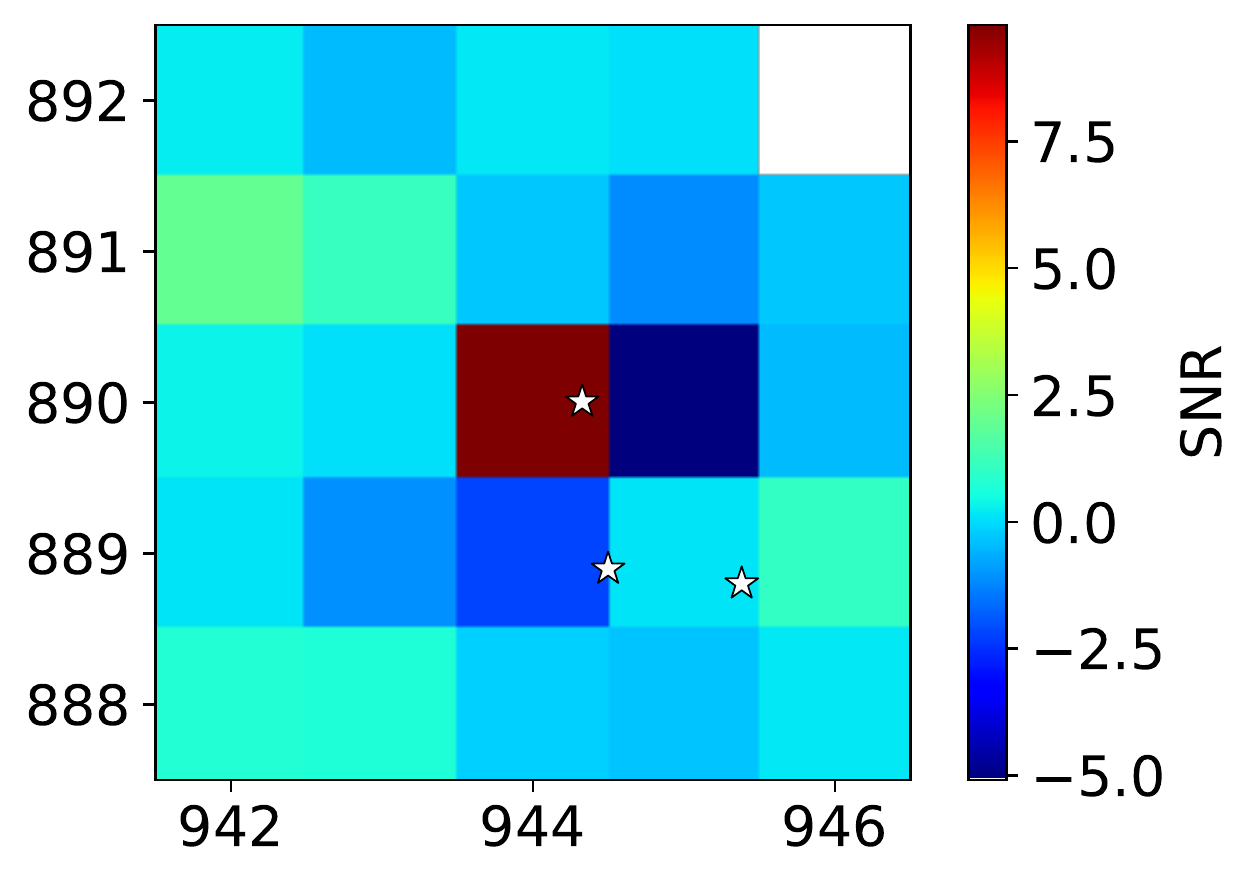}
        \label{fig:SNRQ13}
    \end{subfigure}
\caption{{Mean flux and difference S/N of the new transit candidate on the Kepler pixel map around Kepler-160 (center pixel) for Q5 (a, b), Q9 (c, d), and Q13 (e, f). The positions and Kepler magnitudes of Kepler-160 and two nearby sources are indicated with dots and labels. The difference S/N maps show that the transits are not caused by the nearby sources. }}
\label{fig:lightkurve}
\end{figure}
%**********************************************

{Hence, we used the \texttt{lightkurve} software to generate differential CCD images of the pixel region around Kepler-160 in order to construct per-pixel light curves and measure the S/N of the transit signal for each pixel \citep{2013PASP..125..889B}. In Fig.~\ref{fig:lightkurve} we show the S/N landscape of the new transit candidate in the CCD region around Kepler-160. Visual inspection of the difference images in panels b, d, and f shows that the offset between the source of the new transit candidate signal and Kepler-160 is less than a pixel. The difference images also rule out the possibility that the transit source is due to one of the known stars in the aperture used to extract the light curve (dotted line in Fig.~\ref{fig:pixelmap}). Moreover, Fig.~\ref{fig:lightkurve} rules out the possibility that the source of the transit is due to one of the three stars just outside the aperture. Figure~\ref{fig:lightkurve} does not positively identify the transit signal as being on the target star, but it also does not rule it out. It does confirm, however, that the transits are certainly not caused by any of the nearby sources. We adhere to the ``innocent until proven guilty'' philosophy of the Kepler classification of planet candidates \citep{2016ApJ...822...86M,2018ApJS..235...38T} and identify this transit signal as belonging to a new planet candidate, KOI-456.04.}

We conclude that the three stars just outside the Kepler aperture did not induce the transit candidate signal. {The signal rather comes from within a confusion radius of no more than $3.98\,\arcsec$ around Kepler-160, which is the edge length of one Kepler pixel corresponds to $3.98\,\arcsec$ on the sky.}

\subsubsection{Planet vetting: astrophysical false positives}
\label{sec:FPP}

We set the {\tt maxrad} parameter of \texttt{vespa} equal to $\rho~=~3.98\,\arcsec$ and analyzed the phase-folded transit light curves of Kepler-160\,b and c and of the transit candidate signal. Planets b and c had been statistically validated before \citep{2014ApJ...784...44L,2014ApJ...784...45R}, but no FPPs were published in these studies. \citet{2016ApJ...822...86M}, using \texttt{vespa}, derived FPPs of $3.1~\times~10^{-1}$ for planet b and $1.1~\times~10^{-2}$ for planet c, both of which would not qualify them as statistically validated. We therefore decided to consistently derive the FPPs for all three planets {again}. For planets b and c, we supplied \texttt{vespa} with the data within $\pm$1\,d around the transit mid-point of the phase-folded detrended light curve, which we constructed from the median values for $T_0$ and $P$ as obtained with our MCMC fitting method. For both planets, we searched for secondary eclipses with \texttt{TLS,} and we used our constraints from the nondetections to set the maximum depth for a possible secondary eclipse (the {\tt secthresh} parameter in \texttt{vespa}) to 10 parts per million (ppm). For the new transiting planet candidate, we used $\pm$\,2\,d worth of data around the transit mid-point in the phase-folded light curve and a maximum depth for a possible secondary eclipse of 38\,ppm, using constraints obtained with \texttt{TLS}.

For planets b and c and for the candidate, \texttt{vespa} predicts ${\rm FPP}_1$ values of { $1.95\,{\times}\,10^{-5}$, $6.76\,{\times}\,10^{-2}$, and $8.33\,{\times}\,10^{-2}$}, respectively, where the subscript 1 refers to the assumption of each planet being the only planet around the star. While our FFP$_1$ value for planet c is about a factor of six lower than the value obtained by \citet{2016ApJ...822...86M}, our value for planet b is about four orders of magnitude lower than their value. Part of this discrepancy might lie in the fact that \citet{2016ApJ...822...86M} used DR24 data, whereas we used data from DR25. Although there have been substantial improvements in the Kepler Data Processing Pipeline between DR24 to DR25 \citep{2016AJ....152..158T}, our detailed detrending of any nontransit activity combined with our stringent upper limits on the presence of secondary eclipses (see Sect.~\ref{sec:detrending}) probably also contributed to these improvements of the FPP$_1$ estimates.

For each planet, the corresponding probability of planethood, assuming it were the only transit candidate around Kepler-160, is given as $P_1=1-{\rm FPP}_1$. Taking into account the presence of two additional transit sequences in the light curve of Kepler-160, we can use the empirical evidence of the nonrandomness of transiting exoplanets among Kepler for each of the three transit sequences. Stars that are known to have one transiting planet are more likely to have more transiting planets than a randomly chosen star. In particular, the probability of a planet candidate that is detected around a Kepler star known to have two additional transiting planet candidates can be estimated as \citep{2012ApJ...750..112L}

\begin{equation}\label{eq:fpp2}
P_3 \approx \frac{50 P_1}{50 P_1 + (1-P_1)} \ .
\end{equation}

In the cases of Kepler-160\,b, c, and the new candidate, this ``multiplicity boost'' yields statistically corrected FPP$_3$ values of {$3.90\,{\times}\,10^{-7}$, $1.45\,{\times}\,10^{-3}$, and $1.81\,{\times}\,10^{-3}$}, respectively. The presence of the new nontransiting planet Kepler-160\,d (see Sect.~\ref{sec:non}), however, is not taken into account in these calculations. Strictly speaking, the FPPs of Kepler-160\,b and c and of the new transiting candidate signal are therefore even smaller than our FPP$_3$ values.

All these FPP$_3$ values are significantly below the value of 1\,\% that is commonly used as a threshold for statistical validation \citep{2016ApJ...822...86M}. This means that each of the three transit sequences \textit{could} be formally considered as belonging to a statistically validated planet {with very low astrophysical FPPs due to a blending, grazing, or background eclipsing binary or due to a blend of a transiting planet around an unrelated star near the Kepler CCD location of Kepler-160}. The entire picture, however, is more complicated {because instrumental or statistical false alarms are possible}.

\subsubsection{Planet vetting: instrumental and statistical false alarms}
\label{sec:rolling}

Although the new transit candidate is very unlikely to be an astrophysical false positive, its 378\,d orbital period is suspiciously close to the Kepler heliocentric orbital period of 372\,d, a regime that has been known as a source of the rolling-band artifact \citep{2018AJ....155..210M,2018ApJS..235...38T}. During the primary Kepler mission, the telescope was rolled every 90\,d to keep the solar arrays oriented towards the Sun and the radiator pointed towards deep space \citep{2016ksci.rept....9T}. This maneuver meant that the Kepler target stars moved on and off a given set of Kepler detectors with some periodicity that can result in a transit-like feature in the light curve with a period close to 372\,d after four such rolls. The technical source of the rolling-band effect is in an aliased high-frequency (GHz) noise in the amplifier circuit of the local detector electronics {(\citealt{2010SPIE.7742E..1GK}; Sect.~6.7 in \citealt{2016ksci.rept....1V})}. The exact amplitude and frequency of the resulting artifact depend on the CCD detector channel and the board temperature.

Kepler has 21 science CCD modules and four CCD modules that served as fine-guidance sensors. Each of the 21 science modules has 4 output channels, which yields a total of 84 science channels. We infer from the DV Report of Kepler-160 that this target moved repeatedly over modules/outputs 8/2 (Kepler quarters Q1, Q5, Q9, Q13, Q17), 12/2 (Q2, Q6, Q10, Q14), 18/2 (Q3, Q7, Q11, Q15) and 14/2 (Q4, Q8, Q12, Q16). Most alarmingly, module/output 8/2 has been flagged in the Kepler Instrument Handbook as having a high Moir{\'e} amplitude that may affect the interpretation of the resulting time series \citep[Table 13 in][]{2016ksci.rept....1V}. The three transits that we found occurred in Q5, Q9, and Q13, when Kepler-160 was located on this known rolling-band channel (module/output 8/2).

{We used \texttt{Model-Shift} to evaluate the possibility that this new transit sequence might be caused by rolling-band artifacts alone or in combination with other statistical or instrumental outliers. We fed \texttt{Model-Shift} with our detrended light curve from which we had removed all the transits of planets Kepler-160\,b and c. The resulting light curve thus only contained the three transits of our new candidate. \texttt{Model-Shift} then performed a range of metric tests with the light curve. These metrics have been empirically tested with detections from the Kepler exoplanet search pipeline and simulated data \citep{2018ApJS..235...38T}. \texttt{Model-Shift} computes the multiple event statistic \citep[MES;][]{2002ApJ...564..495J} for the primary transit sequence (Pri), the second most significant sequence (Sec), the third most significant sequence (Ter), and for the most significant positive (or inverted) transit-like feature (Pos).}

%**********************************************
%Fig. 5
\begin{figure}
\centering
\includegraphics[angle= 0,width=1\linewidth]{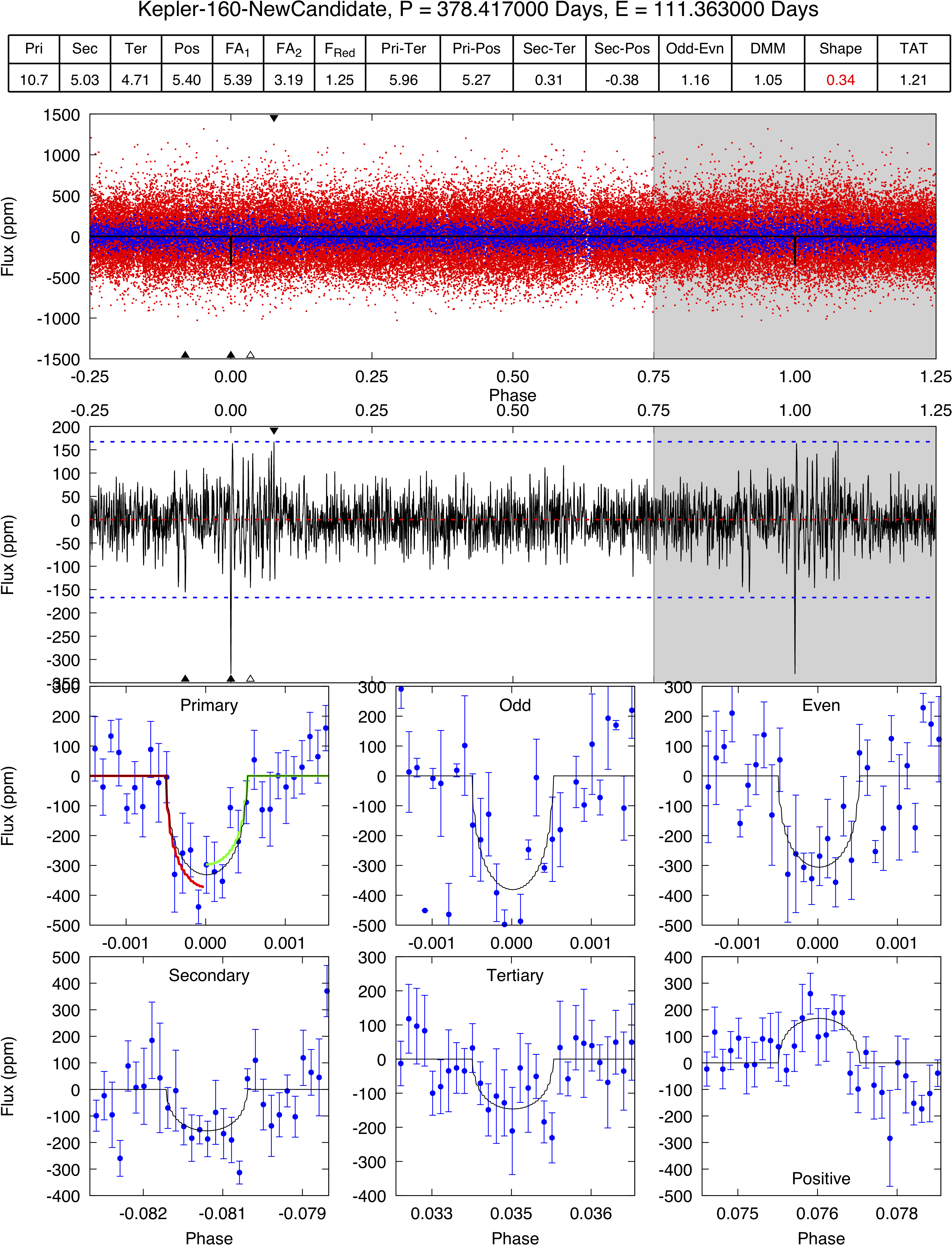}
\caption{{Automated candidate vetting with \texttt{Model-Shift}. The MES values for the Pri, Sec, Ter, and Pos events along with other metrics are listed in the table at the top. The plots show (from top to bottom) the phase-folded light curve without the transits of planets b and c (red: data; blue: binned data; black: model), a zoom into the binned data light curve, and six subpanels illustrating the phase-folded primary, odd, even, secondary, tertiary, and positive events.}}
\label{fig:ModelShift}
\end{figure}
%**********************************************

%**********************************************
%Fig. 6
\begin{figure}
\centering
\includegraphics[angle= 0,width=1\linewidth]{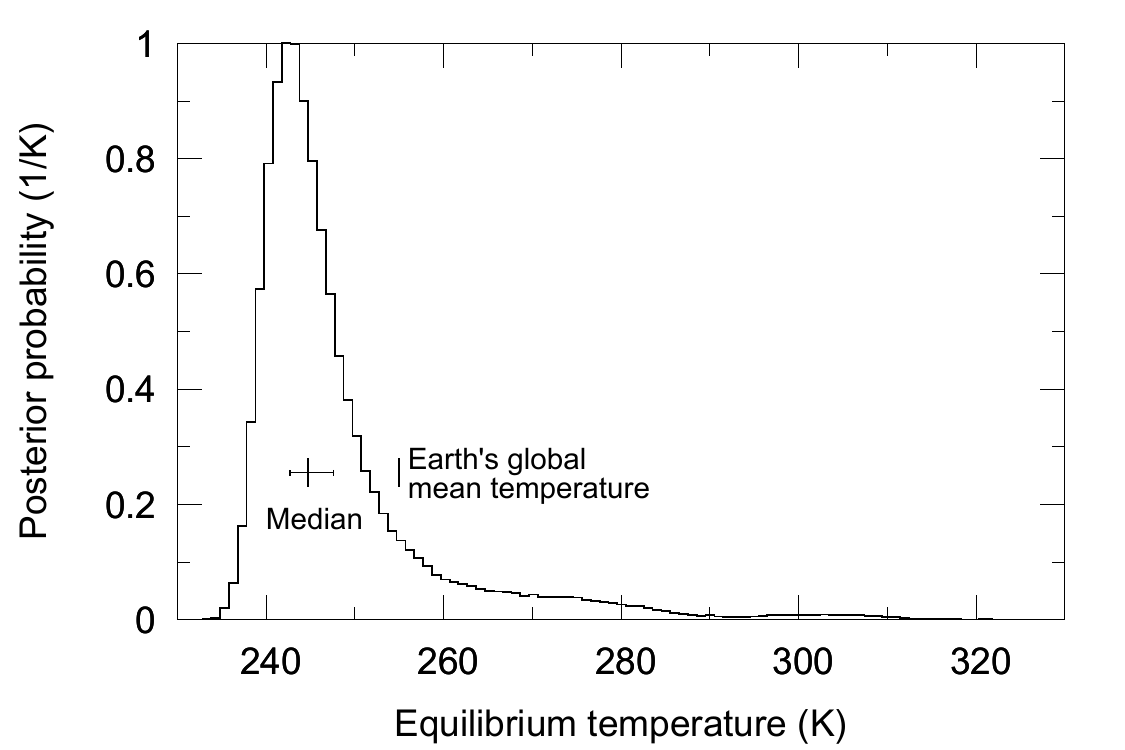}
\caption{Posterior distribution of the surface temperature of the new transiting candidate as derived from 150,000 MCMC samplings of the Kepler light curve of Kepler-160. We assumed an Earth-like Bond albedo of 0.3. The median value of 244.8\,(+2.1, -2.9)\,K is indicated with a vertical marker, as is the Earth's effective global mean surface temperature as calculated from its top-of-the atmosphere solar insolation, that is, neglecting the atmospheric greenhouse effect.}
\label{fig:temperature}
\end{figure}
%**********************************************

%**********************************************
%Fig. 7
\begin{figure*}
\centering
\includegraphics[angle= 0,width=0.495\linewidth]{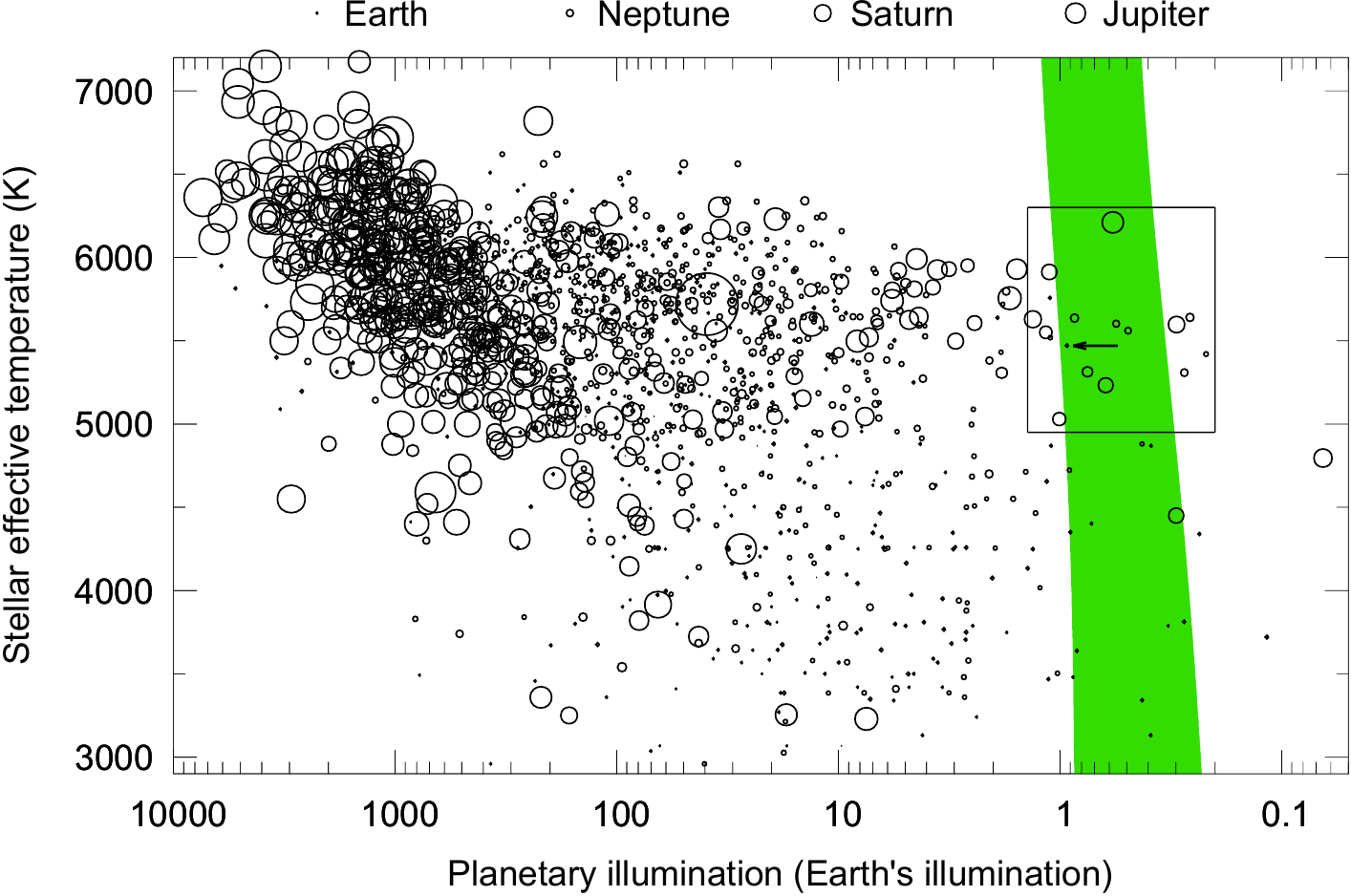}
\includegraphics[angle= 0,width=0.495\linewidth]{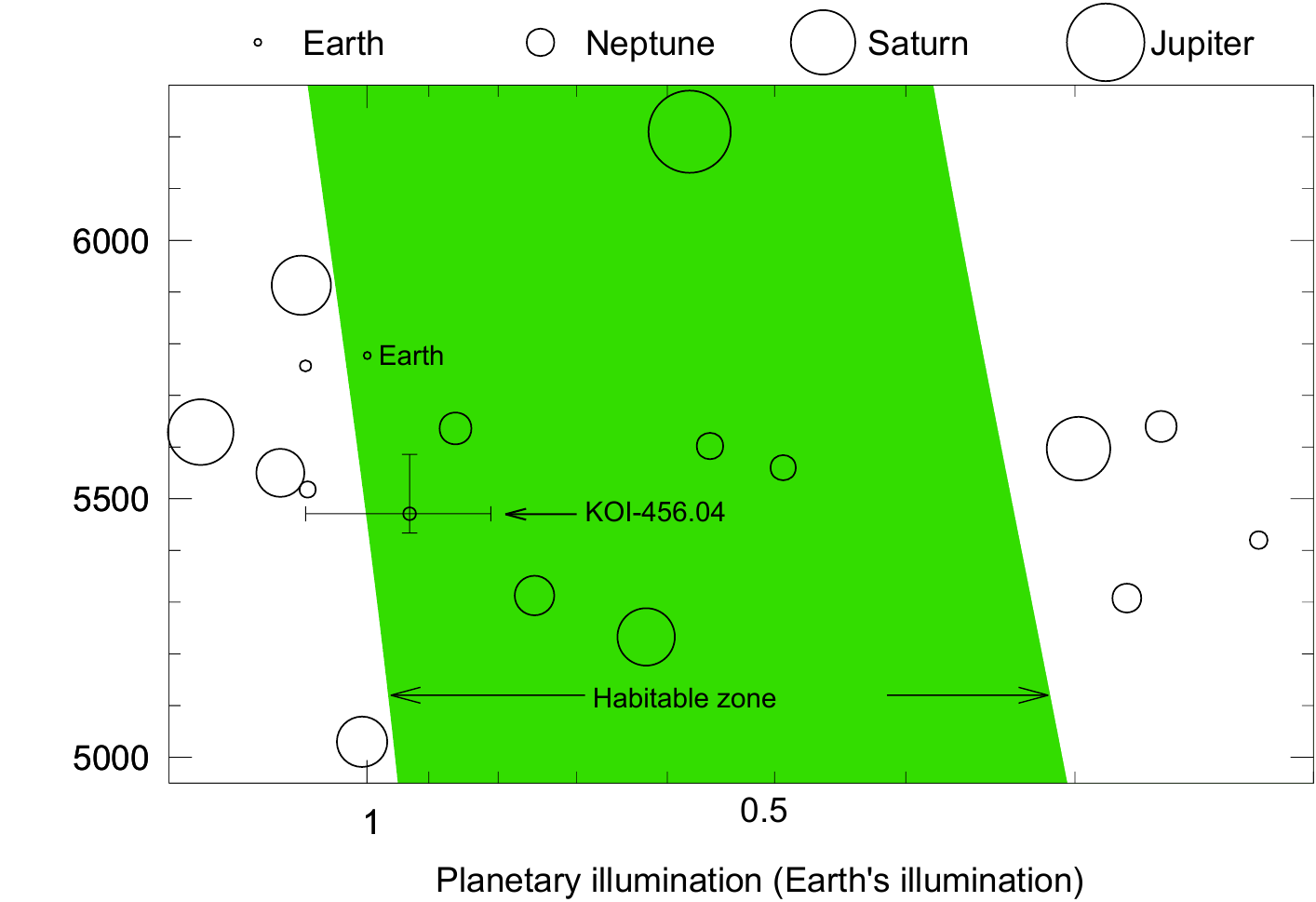}
\caption{New transiting candidate {KOI-456.04} in the context of the known extrasolar planet population. The position of the open circles indicates the stellar flux received at the top of the atmosphere and the effective temperature of the host star of all known transiting planets as listed in the Exoplanet Encyclopaedia (\href{http://exoplanets.eu}{http://exoplanets.eu}) on 29 April 2020. The sizes of the circles scale with the radii of the planets. The green shaded area denotes the habitable zone within the limits of the maximum greenhouse effect (right border) and the runaway greenhouse effect (left border). For details, see Sect.~\ref{sec:HZ}. \textbf{(a)} The rectangle outlines the margins displayed in panel (b), and the arrow points at the new transiting candidate. \textbf{(b)} The Earth's position and size are indicated for reference. The new candidate is labeled with an arrow, and error bars refer to the 1\,$\sigma$ confidence intervals from our MCMC fits.
%The estimated probability of planet $d$ being inside the habitable zone is XX\,\%.
Circles are to scale within each panel but not across panels.}
\label{fig:HZ}
\end{figure*}
%**********************************************

{Figure~\ref{fig:ModelShift} is an automated output file from \texttt{Model-Shift}. The table at the top of Fig.~\ref{fig:ModelShift} contains an overview of the Pri (10.7), Sec (5.03), Ter (4.71), and Pos (5.40) metrics. These values need to be compared to the FA$_1$ value of 5.39, which is the $3\,\sigma$ threshold at which any event is considered to be significant when the measured systematic noise level is taken into account. Although the primary signal is highly significant, the multitude of near-$3\,\sigma$ features suggests that the rolling-band phenomenon is in fact present in the data. This indicates weaker features in the light curve that have a shape similar to the planet candidate transit signal with almost half the strength (see the bottom panels in Fig.~\ref{fig:ModelShift}). While our planet candidate is significantly stronger, it is still possible, though unlikely, that some combination of rolling bands and other noise is causing our candidate signal. As for the Shape metric in the table at the top of Fig.~\ref{fig:ModelShift}, a value near 0 would suggest a transit-like signal and a value near 0.5 a sinusoidal or heartbeat-like signal \citep[see][]{2017ksci.rept...15C}. Our value of 0.336 is just slightly above an arbitrary threshold of 0.3, beyond which the entry is highlighted in red. Because our \texttt{TLS} transit-search algorithm searches for transit-like events with the well-known signal detection efficiency metric \citep[SDE;][]{2002A&A...391..369K,2019A&A...623A..39H}, we do not consider this flag as critical.}

{We used the Pri MES value of 10.7 and the orbital period of 378\,d to infer a ${\sim}85\,\%$ reliability of the candidate against instrumental artifacts using Fig.~22 of \citet{2019arXiv190603575B}. We conclude that this newly discovered transit signal passes the astrophysical vetting (Sect.~\ref{sec:FPP}) and that it has a reasonable reliability against instrumental artifacts. Formally speaking, it remains an (unvalidated) planet candidate, however, because validation would require both FPP$_3<1\,\%$ and reliability against instrumental artifacts $>99\,\%$.}

\subsubsection{Characterization of the transiting habitable zone candidate}
\label{sec:HZ}

%**********************************************
%Fig. 8
\begin{figure}
\centering
\includegraphics[angle= 0,width=1\linewidth]{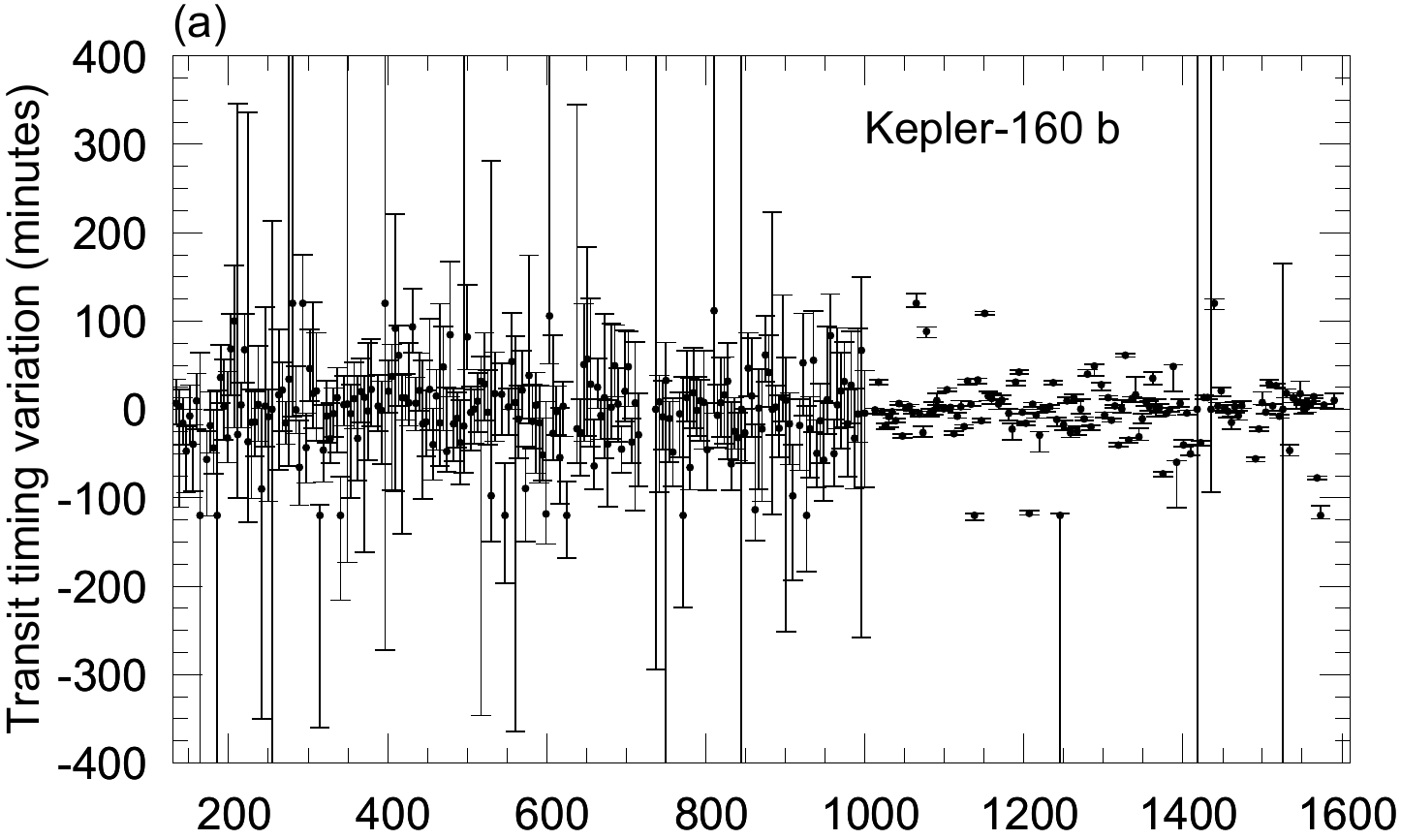}\\
\vspace{0.3cm}
\includegraphics[angle= 0,width=1\linewidth]{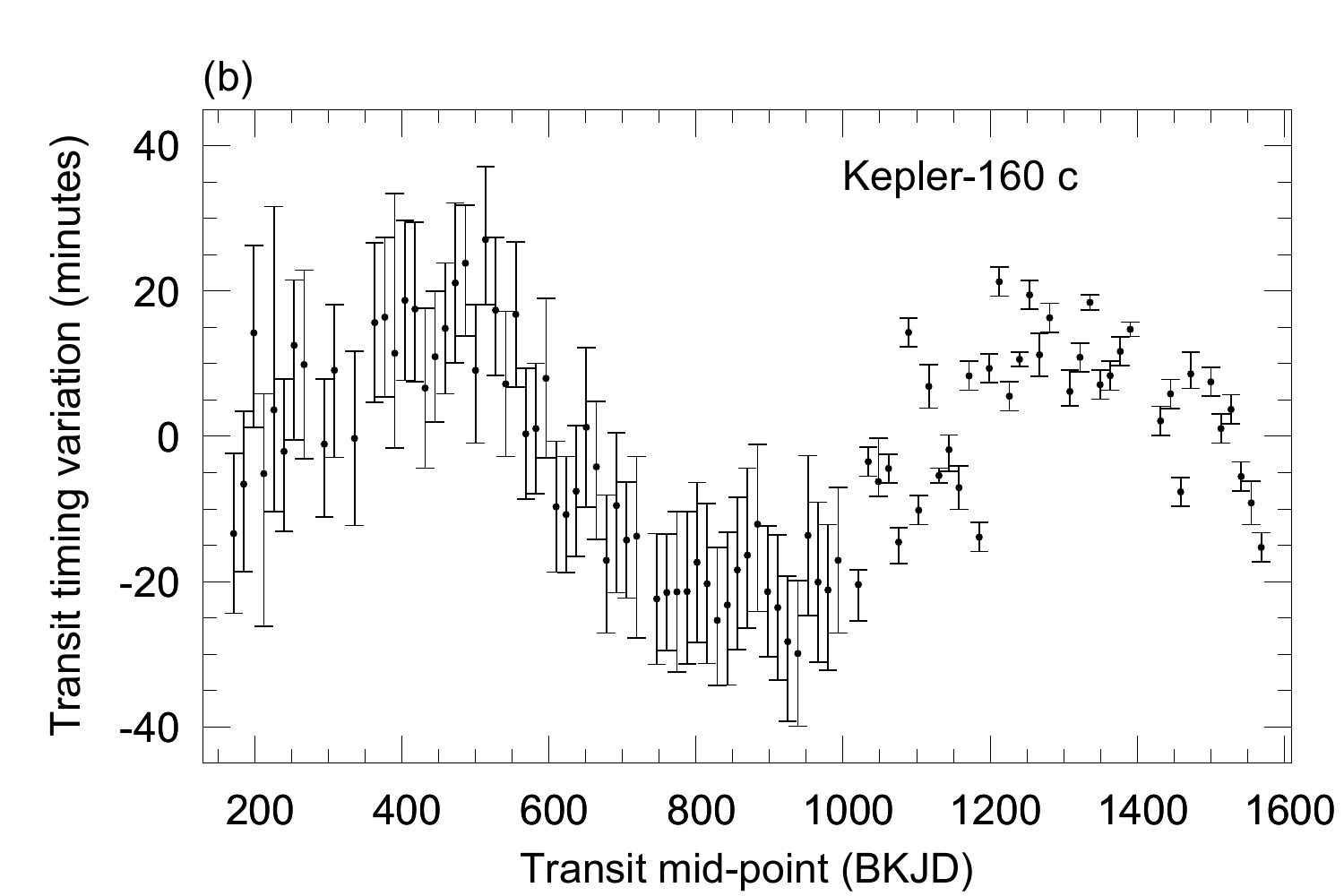}
\caption{\textbf{(a)} TTVs of Kepler-160\,b and \textbf{(b)} TTVs of Kepler-160\,c as derived from our one-planet transit fits to the two transit sequences.}
\label{fig:TTVs}
\end{figure}
%**********************************************

%**********************************************
%Fig. 9
\begin{figure}
\centering
\includegraphics[angle= 0,width=1\linewidth]{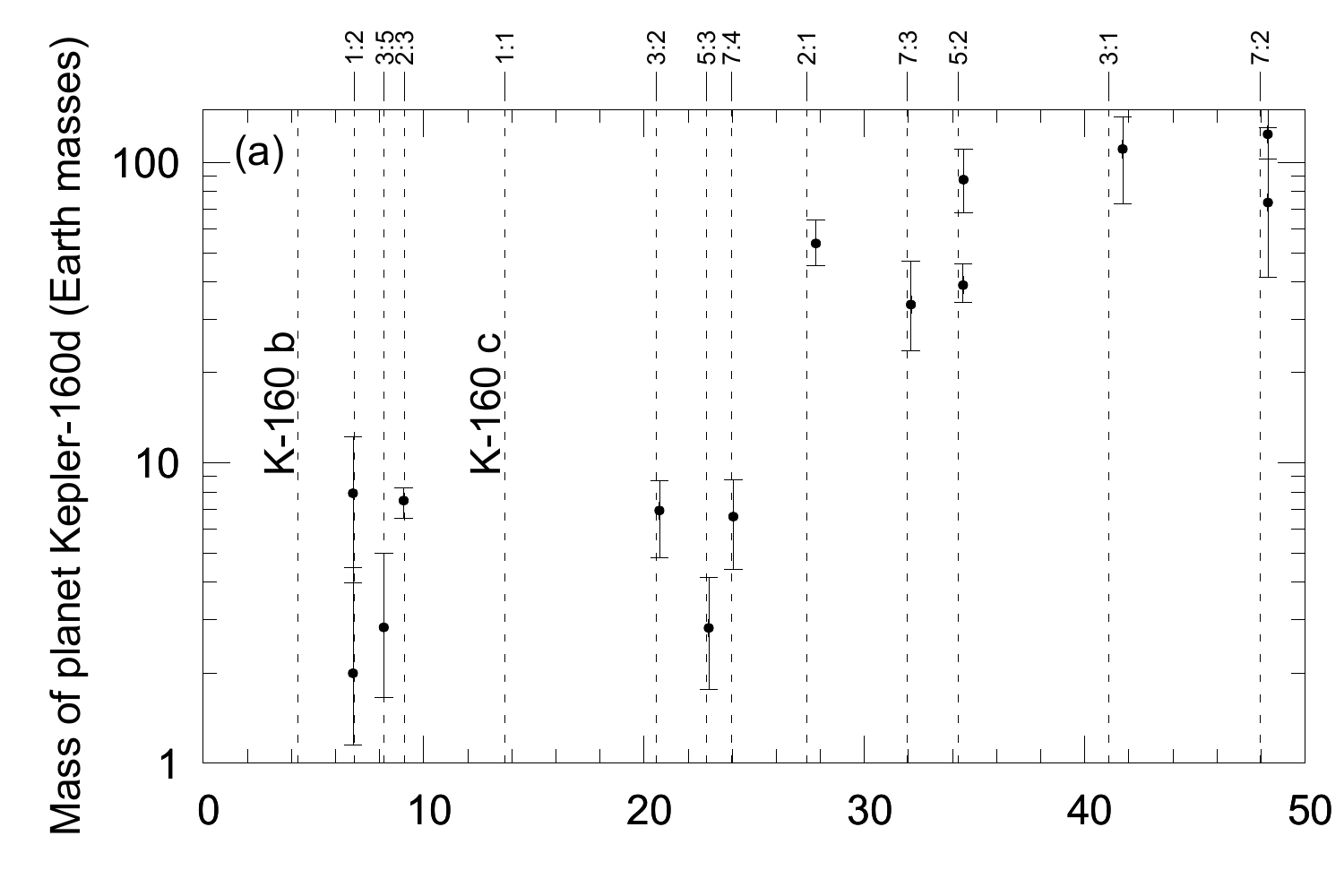}\\
\vspace{0.4cm}
\includegraphics[angle= 0,width=1\linewidth]{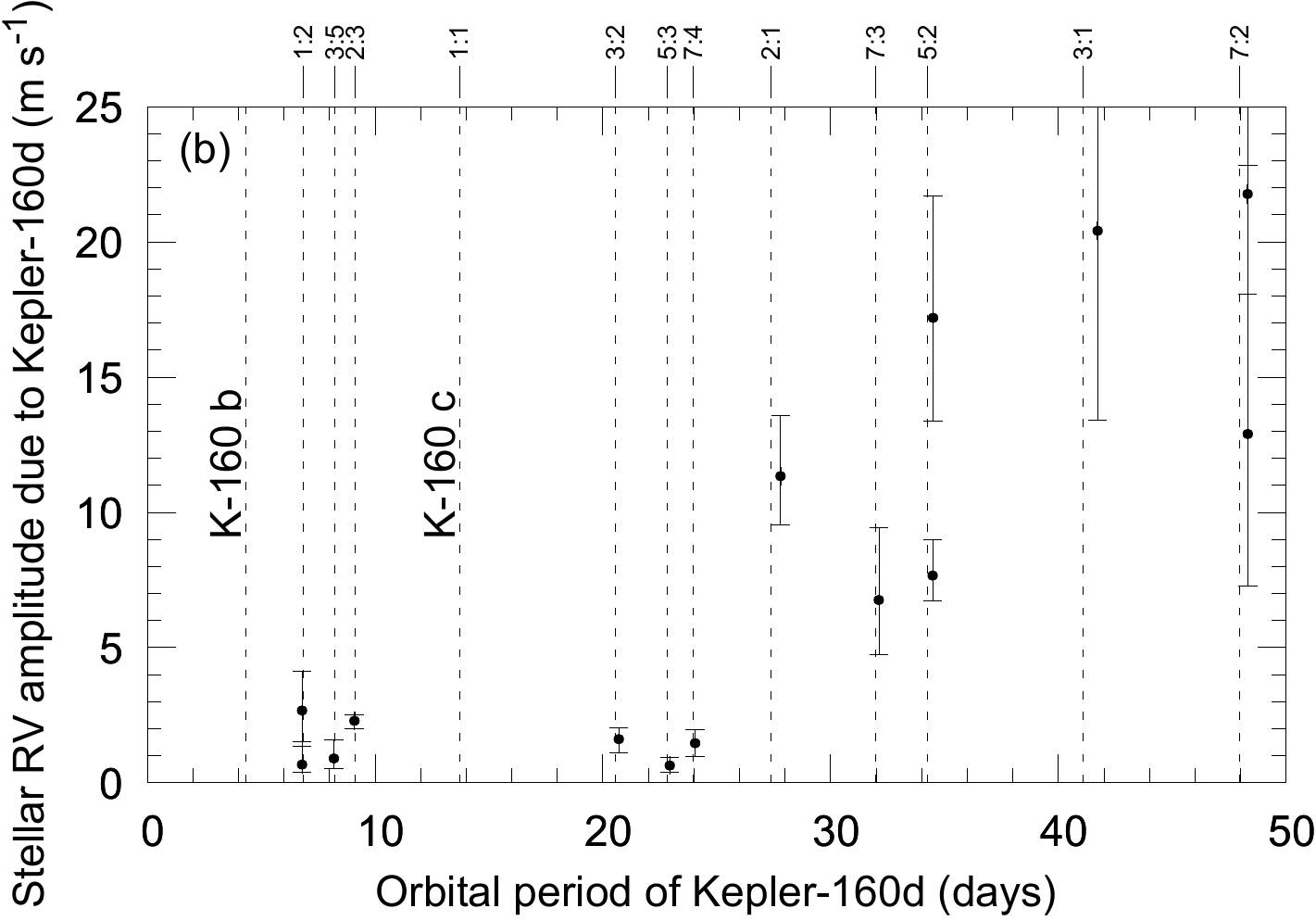}
\caption{Results of our MCMC fitting procedure. \textbf{(a)} Expected mass of the nontransiting planet Kepler-160\,d as a function of its orbital period. Dots signify mean values, and error bars illustrate the posterior distribution with 34\,\% of the sample contained in both positive and negative directions from the mean, respectively. The scaling of the ordinate is logarithmic. The orbital periods of Kepler-160\,b and c and the orbital MMRs with planet c that we tested are indicated with vertical dashed lines (see labels). Two solutions with very similar probabilities exist for both the 5:2 and 7:2 MMRs. \textbf{(b)} Expected RV amplitude of the host star due to the orbital motion of Kepler-160\,d as a function of $P$.}
\label{fig:TTVsRVs}
\end{figure}
%**********************************************

We find that the stellar radiative energy flux of the new transit candidate would be $0.93_{-0.12}^{+0.18}$ times the insolation at the top of the Earth's atmosphere. An Earth-like Bond albedo of 0.3 would result in a globally averaged surface temperature of about $244.8_{-2.9}^{+2.1}$\,K (see Fig.~\ref{fig:temperature}). This value is just $10^\circ$C lower than the nominal value of 255\,K ($-18^\circ$C) for Earth computed from its net absorption of the top-of-the atmosphere solar flux in thermal equilibrium and in absence of an atmosphere. These calculations neglect the additional heating of the greenhouse effect, for instance, from water vapor (H$_2$O), carbon dioxide (CO$_2$), methane (CH$_4$), nitrous oxide (N$_2$O), ozone (O$_3$), and other greenhouse gases. On Earth, the greenhouse effect amounts to about $+33^\circ$C. If the transit candidate signal belongs to a genuine planet and if this planet has an atmosphere that provides an Earth-like greenhouse effect, then the globally averaged surface temperature is around $+5^\circ$C.

In Fig.~\ref{fig:HZ}(a) we show the distribution of all known transiting exoplanets in the context of the conservative stellar habitable zone \citep[HZ, green shaded area;][]{1993Icar..101..108K} around their individual stars, 3038 planets in total. Most transiting exoplanets are likely gas-dominated giant planets, their median (mean) radius is $2.3\,R_\oplus$ ($4.3\,R_\oplus$) or 0.6 (1.1) Neptune radii. The hot limit of the conservative HZ is given by the runaway greenhouse effect derived from one-dimensional radiative-convective cloud-free climate models of water-rich, rocky planets \citep{2013ApJ...765..131K}. The cold limit is imposed by the maximum greenhouse effect that could be provided by a CO$_2$-rich atmosphere at the limit of condensation. Only a few transiting planets are known to orbit their star in the habitable zone.

Figure~\ref{fig:HZ}(b) shows a zoom into the region of planets with Earth-like insolation around Sun-like stars. In this region of the conservative HZ, the new transiting candidate (denoted with an arrow) would be the smallest planet by far and the only super-Earth with a radius smaller than two Earth radii.

\subsection{Nontransiting planet Kepler-160\,d}
\label{sec:non}

%**********************************************
%Fig. 10
\begin{figure}
\centering
\includegraphics[angle= 0,width=1.02\linewidth]{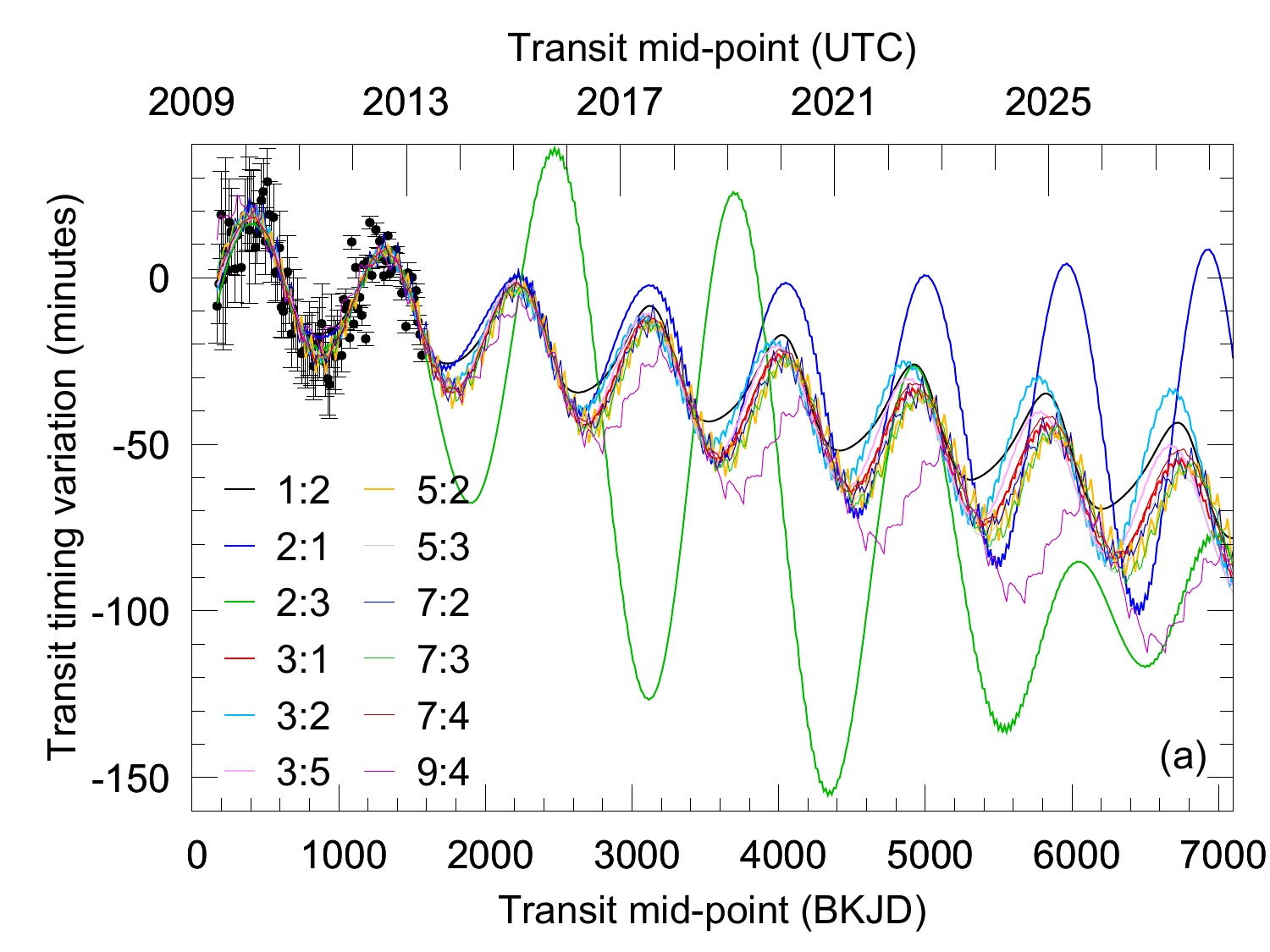}\\
\vspace{0.35cm}
\includegraphics[angle= 0,width=0.99\linewidth]{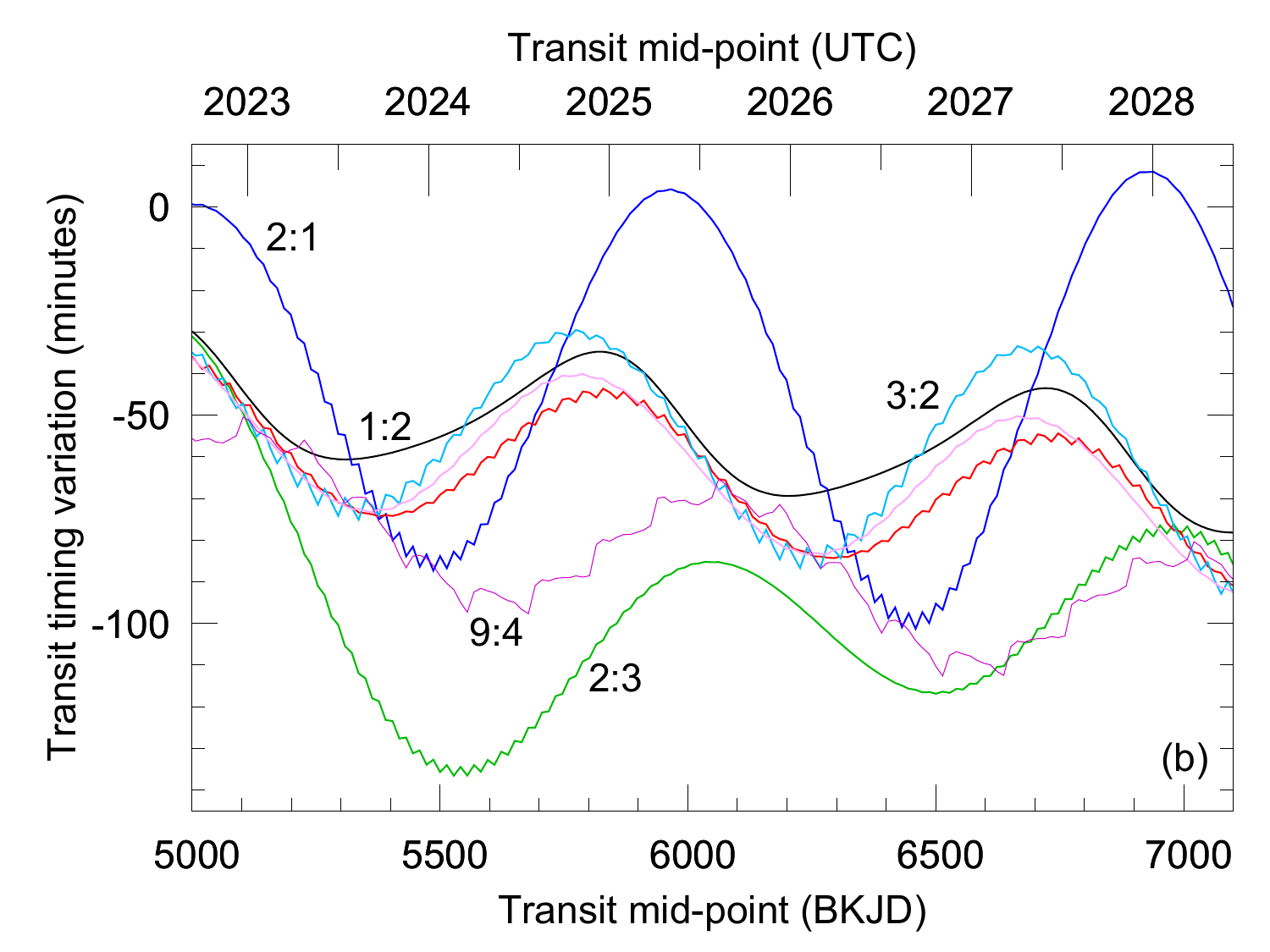}
\caption{Predicted TTVs of Kepler-160\,c for various MMRs of the nontransiting planet Kepler-160\,d. Times of the transit mid-points are given along the primary x-axis in units of the BKJD and on the secondary x-axis in units of the Coordinated Universal Time (UTC). For reference, BKJD = 7000\,d corresponds to 2 March 2028. \textbf{(a)} Predicted TTV curves for all resonances shown in Fig.~\ref{fig:TTVsRVs}. The color-coding of the curves is described in the legend. \textbf{(b)} Zoom into the final $\sim2000$\,d of our simulations. For clarity, the curves are restricted to some of the most diverging resonances (see labels) and the 5:2, 5:3, 7:2, 7:3, and 7:4 MMRs have been excluded.}
\label{fig:TTVprediction}
\end{figure}
%**********************************************

In Fig.~\ref{fig:TTVs} we show the TTVs of Kepler-160\,b and c as derived with our fittings of a one-planet transit model to each individual transit.\footnote{In Fig.~\ref{fig:TTVs}(b) we do not show four TTV values with extremely large error bars of up to half a day (or about 700\,min), which were caused by a low number of in-transit data points. These transits occurred around BKJDs of 281.1\,d, 349.6\,d, 1404.4\,d, and 1418.1\,d, respectively.} The decrease in the width of the error bars after BKJD~=~1000\,d is due to the availability of short-cadence data from that point on, whereas earlier measurements were taken in long-cadence mode. The TTVs of planet b do not show any significant periodicity, but the periodicity of the TTVs of planet c is very pronounced. The TTV amplitude of approximately 20\,min and the TTV period of about 879\,d is in good agreement with earlier studies by \citet{2016ApJS..225....9H} and \citet{2018ApJS..234....9O}.

We also inspected the Lomb-Scargle periodograms of the TTV series of both planets (not shown). For Kepler-160\,b we find no clear signal, and for planet c we find that the 879\,d period is the strongest periodic signal.

In Fig.~\ref{fig:TTVsRVs} we visualize the results of our three-body TTV modelling including the star Kepler-160, planet~c, and one perturber near a low number period commensurability to planet~c. Panel~(a) shows the posterior distribution of the mass of the planet ($M_{\rm d}$) that would act as a gravitational perturber to cause the observed TTVs of Kepler-160\,c. We tested planets near various period commensurabilities (see labels at the top) and find that $M_{\rm d}$ ranges between about 1 and $100\,M_\oplus$ within the error bars, details depending on the period of the planet. Our results show that various solutions for a nontransiting planet exist and that more data are required before a unique solution can be determined. At this point, our purpose therefore is not a full study of the parameter space, and we accordingly restrict our analysis to the selection of period commensurabilities shown in Fig.~\ref{fig:TTVsRVs}. Some periods have two mass solutions, where the \texttt{emcee} optimization resulted into two solutions with similar likelihood. Generally speaking, a planet interior to the orbit of Kepler-160\,c could have $M_{\rm d}\,\sim\,1\,M_\oplus$, whereas a planet in an outer orbit might be much more massive. These mass estimates of the nontransiting disturber are sensitive to its orbital inclination with respect to Kepler-160\,c. Our results are only formally valid for a hypothetical perturber transiting very near our line of sight. That said, the order of magnitude of $M_{\rm d}$ would still be similar if the mutual inclination of planet c and the perturber were small.

We also tested the hypothesis that the 879\,d period in the TTV series of Kepler-160\,c is caused by the same planet that is also responsible for the new transit candidate signal detailed above. These attempts, however, did not produce a consistent solution, which prevents us from associating the new transit candidate with the observed TTVs and thus confirming it. In summary, all our simulations suggest that the perturber responsible for the observed TTVs of Kepler-160\,c is (a) deep in the planetary mass range and (b) cannot be associated with either the known transiting planet Kepler-160\,b or the new transiting candidate. We therefore conclude that there is another nontransiting planet in the system, which we label Kepler-160\,d, and which might be described by one of the solutions shown in Fig.~\ref{fig:TTVsRVs}.

Figure~\ref{fig:TTVsRVs}(b) predicts the stellar radial velocity (RV) amplitude that would result from the nontransiting planet Kepler-160\,d {in a near line-of-sight orbit}. These values range between about 1 and $20\,{\rm m\,s}^{-1}$, which could make Kepler-160\,d detectable {using} ground-based stellar spectroscopy.

For the mean $P$ and $M_{\rm d}$ values shown in Fig.~\ref{fig:TTVsRVs} we calculated the transit times of Kepler-160\,c over roughly 7000\,d, that is, until about March 2028, with our TTV model, the results of which are shown in Fig.~\ref{fig:TTVprediction}. Panel (a) shows the TTVs of all configurations that we considered and panel (b) shows a zoom into the final 2000\,d of the predictions for a selection of configurations. We find that many of our TTV predictions for Kepler-160\,c are very similar to the 1:2 MMR (solid black line). Most notably, the TTV divergence of the 2:1 (blue line), 2:3 (green line), and 9:4 (purple line) MMRs exceeds the typical error of the Kepler short-cadence data for a substantial part of our simulations. These configurations might therefore most easily be tested with follow-up transit observations of Kepler-160\,c. Alternatively, stellar RVs might be {used}.

\section{Discussion}
\label{sec:discussion}

\subsection{Follow-up}

% Table 2
\begin{table*}
%\scriptsize
\centering
\caption{Mid-times of the upcoming possible transits of the new transit candidate {(KOI-456.04)} around Kepler-160.}
\def\arraystretch{1.5}
\label{tab:transits}
\centering
\begin{tabular}{cccc}
\hline
Transit mid-point (BKJD) & Transit mid-point (UTC) & Begin of ingress (UTC) & End of egress (UTC)\\
\hline
4273.96$^{+0.68}_{-0.69}$\,d & 14.09.2020, 11:06$^{+16\rm\,h}_{-17\rm\,h}$ & 14.09.2020, 04:40$\pm$\,16\,h & 14.09.2020, 17:32$\pm$\,17\,h\\
4652.38$^{+0.75}_{-0.76}$\,d & 27.09.2021, 21:10$\pm$\,18\,h                & 27.09.2021, 14:44$\pm$\,18\,h & 28.09.2021, 03:35$^{+18\rm\,h}_{-19\rm\,h}$\\
5030.80$^{+0.83}_{-0.84}$\,d & 11.10.2022, 07:14$\pm$\,20\,h & 11.10.2022, 00:47$\pm$20\,h & 11.10.2022, 13:39$^{+20\rm\,h}_{-21\rm\,h}$\\
5409.22$^{+0.90}_{-0.91}$\,d & 24.10.2023, 17:17$\pm$\,22\,h & 24.10.2023, 10:51$^{+21\rm\,h}_{-22\rm\,h}$ & 24.10.2023, 23:42$\pm$\,22\,h\\
$5787.64_{-0.97}^{+0.99}$\,d &
06.11.2024, 03:21$_{-23}^{+24}$\,h &
05.11.2024, 20:54$_{-23}^{+24}$\,h &
06.11.2024, 09:46 $\pm24$\,h \\
$6166.06_{-1.05}^{+1.07}$\,d &
19.11.2025, 13:24 $_{-25}^{+26}$\,h &
19.11.2025, 06:58 $_{-25}^{+26}$\,h &
19.11.2025, 19:49 $_{-25}^{+26}$\,h \\
$6544.48_{-1.12}^{+1.15}$\,d &
02.12.2026, 23:28 $_{-27}^{+28}$\,h &
02.12.2026, 17:02 $\pm27$\,h &
03.12.2026, 05:53 $_{-27}^{+28}$\,h \\
$6922.89_{-1.20}^{+1.22}$\,d &
16.12.2027, 09:31 $\pm29$\,h &
16.12.2027, 03:05 $\pm29$\,h &
16.12.2027, 15:56 $_{-29}^{+30}$\,h \\
$7301.31_{-1.27}^{+1.30}$\,d &
28.12.2028, 19:34 $\pm31$\,h &
28.12.2028, 13:08 $_{-30}^{+31}$\,h &
29.12.2028, 02:00 $_{-31}^{+32}$\,h \\
\hline
\end{tabular}\\
\tablefoot{Values are the median and 3\,$\sigma$ uncertainties from the posterior samples of our MCMC fitting procedure to the Kepler PDCSAP detrended light curve. All times are corrected for light travel time between Earth and the Solar System barycenter.}
\end{table*}

The current state of technology and the {available} astronomical facilities make it extremely challenging to validate or reject {the new transit candidate KOI-456.04}. In principle, such a validation could be achieved through additional transit observations, stellar RV measurements, or by measuring TTVs of the transiting planets that can be explained by a planet with an orbital period of 378\,d.

The resulting RV amplitude of its solar-mass host star \citep{2017AJ....154..108J} can be estimated to range between about 0.3\,m\,s$^{-1}$ for an ocean world and 1.2\,m\,s$^{-1}$ for a planet with Earth-like composition. Although these RV amplitudes might be accessible with the Echelle Spectrograph for Rocky Exoplanets and Stable Spectroscopic Observations (ESPRESSO) \citep{2010SPIE.7735E..0FP} at the Very Large Telescope, Kepler-160 itself is not observable from Chile. This transit candidate could nevertheless be a suitable target to revisit with future spectrographs on the Northern Hemisphere constructed for the RV detection of super-Earths in Earth-like orbits around Sun-like stars. This would be an important stepping stone to confirm even lower-mass Earth-mass planets that will need to be followed-up after their expected discovery with the PLATO mission \citep{2014ExA....38..249R}, with an expected launch in 2026.

If Kepler-160 were reobserved with PLATO, then it would not nominally be in any of the stellar samples (P1 to P5), their limiting apparent magnitudes being $m_V = 11$ and 13, respectively.\footnote{PLATO Definition Study Report: \href{https://sci.esa.int/s/8rPyPew}{https://sci.esa.int/s/8rPyPew}} That said, dedicated imagette analysis within the guest observer program might be possible. The potential for a recovery of the candidate signal {of KOI-456.04} will depend on the actual location of the star on the PLATO field of view. PLATO will have 24 normal cameras (plus 2 fast cameras), and any star will be observed by either 6, 12, 18 or 24 cameras synchronously. The PLATO performance and noise behavior is expected to be comparable to that of Kepler for the stars that will be observed with all 24 cameras, that is to say, in the center of the field of view. The random noise of an $m_V = 11$ star is expected to be 34\,ppm\,hr$^{-1}$ for a combination of data from 24 cameras. In the optimal case of a coverage by all 24 cameras, the flux of Kepler-160 would be $10^{(11-14.6)/2.5} = 3.6$\,\% compared to that from that from an $m_V = 11$ star. This means that the random noise will be 34\,ppm\,hr$^{-1}$/$\sqrt{3.6\,\%} \sim 179$\,ppm\,hr$^{-1}$, which is slightly more than half of the measured transit depth (see Sect.~\ref{sec:app_detrending}). Such a configuration or, equivalently, the coverage with 12 cameras and monitoring of 2 transits, might therefore in principle be able to marginally recover the transits of {KOI-456.04} if it genuinely is a planet.

In Table~\ref{tab:transits} we predict the next nine transits of this transit candidate based on our MCMC sampling of the entire light curve. We find very long transit durations (about 13\,hr) for this candidate, which allows many measurements during a single telescope visit. This fact compensates to some extent for the relative faintness ($G=14.6160, J=13.460, H=13.101, K=13.052$) \citep{2003yCat.2246....0C,2018yCat.1345....0G} of this star. The faintness is typical for the Kepler sample. We also find that the uncertainty in the transit times increases by about 2\,hr between transits, that is, at a rate of roughly 2\,hr/yr. Consequently, observations of one of the upcoming putative transits listed in Table~\ref{tab:transits} will be very important to secure or reject the planet interpretation of this signal.

\subsection{Planet interpretation of the new transit candidate}

The radius of $1.91^{+0.17}_{-0.14}\,R_\oplus$ of the transit candidate suggests that it might be surrounded by a voluminous layer of volatiles such as H or He \citep{2015ApJ...801...41R}.
%It may thus be beyond the} transition regime between rock-/water-dominated planets with thin atmospheres and gas-dominated planets with small cores.
Although direct mass measurements of the candidate are currently not available and the mean density is unknown, planetary evolution tracks \citep{2014ApJ...792....1L} suggest that any potential hydrogen or helium atmosphere would contribute less than about 0.75\,\% to its total planetary mass.
%According to a simple mass-radius scaling relation derived from solar system observations\cite{2011ApJS..197....8L}, the radius of $1.91\,R_\oplus$ is suggestive of a planet with a mass of about 3.8 Earth masses.
Planetary structure models \citep{2007ApJ...659.1661F} suggest that for an Earth-like (or, alternatively, a purely rocky) composition the planetary mass would be about or $13.5\,M_\oplus$ (or $9.9\,M_\oplus$). If the transit candidate, however, were an ocean world composed of equal amounts of rock and water, then its mass would only be about $3.5\,M_\oplus$ for the measured radius.

Although these properties do not seem to make it very Earth-like, its putative super-Earth mass and estimated age of $8.9_{-1.7}^{+4.2}$ billion years \citep{2017AJ....154..108J} could nevertheless 
make it a superhabitable world \citep{2014AsBio..14...50H}, potentially with even better prospects for life than truly Earth-like planets.

\subsection{Alternative stellar parameterization}
\label{sec:alternative}

The $T_{\rm eff}$ value of $\,5471_{-37}^{+115}$\,K from the Gaia DR2 that we used for the purpose of consistency throughout our analysis is somewhat lower than the values of $5616_{-64}^{+65}$\,K determined from HIRES high-resolution spectroscopy \citep{2017AJ....154..108J} and $5538_{-75}^{+75}$\,K derived from RCSpec mid-resolution spectroscopy \citep{2013ApJ...771..107E}. The stellar radius of $1.118_{-0.045}^{+0.015}\,R_\odot$ from Gaia DR2 is well within the uncertainties of $R_{\rm s}~=~1.11_{-0.11}^{+0.13}\,R_\odot$ as measured based on HIRES spectroscopy \citep{2017AJ....154..108J}, although it is somewhat larger than the value of 1.04\,($\pm$\,0.05)\,$R_\odot$ estimated via RCSpec spectroscopy \citep{2013ApJ...771..107E}. The stellar effective temperature of 5857\,($\pm$\,200)\,K and radius of 0.884\,($\pm$\,0.430)\,$R_\odot$ given in the Kepler Input Catalog \citep{2011AJ....142..112B,2017ApJS..229...30M} (available at the NASA Exoplanet Archive)\footnote{\href{https://exoplanetarchive.ipac.caltech.edu}{https://exoplanetarchive.ipac.caltech.edu}} were derived using five-band photometry without additional stellar spectroscopy or model-independent parallax measurements. Instead, the stellar extinction between Earth and the source was estimated and used as a free parameter, which is why we consider these values as less reliable.

The Gaia DR2 value of the stellar radius that we used is the highest of all the sources discussed above. As a consequence, and because $R_{\rm p}$ scales linearly with $R_{\rm s}$, we derive relatively large planetary radii. If we were to use best-fitting stellar radii of 1.11\,$R_\odot$ \citep{2017AJ....154..108J}, 1.04\,$R_\odot$ \citep{2013ApJ...771..107E}, or 0.884\,$R_\odot$ \citep{2011AJ....142..112B,2017ApJS..229...30M}, for comparison, the resulting planetary radius for candidate transit signal would be 1.89\,$R_\oplus$, 1.77\,$R_\oplus$, or 1.50\,$R_\oplus$, respectively. This shows that although the different literature values for the stellar radius all agree within about two standard deviations, the effect on the planetary radius might have dramatic implications for the composition of the planet and its nature as a super-Earth or a mini-Neptune \citep{2015ApJ...801...41R}.

\subsection{Coplanarity of the nontransiting planet Kepler-160\,d}
\label{sec:coplanarity}

In our MCMC optimization procedure of the TTVs of Kepler-160\,c (Sect.~\ref{sec:TTVs}) we assumed an orbital inclination of $90^\circ$ of the test planet, that is to say, a line-of-sight arrangement of its orbit. Strictly speaking, this assumption contradicts our nondetection of transits from our newly discovered planet, Kepler-160\,d, which our analysis suggests to have a mass (and we expect thus a radius) in the super-Earth or mini-Neptune regime of maybe between $2\,R_\oplus$ and $4\,R_\oplus$. If the orbital inclination of Kepler-160\,d were truly $90^\circ$, then we would almost certainly have detected its transits.

A short trigonometric exercise shows that the critical orbital inclination for the disk center of Kepler-160\,d to perform a grazing transit is $i_{\rm g}=\arccos(R_{\rm s}/a)$. The possible orbital periods of Kepler-160\,d between about 5\,d and 50\,d, as suggested by our TTV fitting procedure (see Fig.~\ref{fig:TTVsRVs}), correspond to orbital semimajor axes between about 0.06\,AU and 0.27\,AU. The resulting values for $i_{\rm g}$ then range between $84^\circ.8$ in the innermost orbits and $88^\circ.9$ in the outer orbits.

Our results for the mass and semimajor axis of the test planet will only differ by a few percent if its actual inclination differs by only a few degrees from our line-of-sight assumption. We argue that it is very plausible for Kepler-160\,d to be on a near-coplanar orbit with planet c given that planet c is essentially coplanar with planet b, given the near-coplanarity of the Solar System planets, and given the many detections of near-coplanar multiplanet transiting systems from Kepler \citep{2014ApJ...784...45R}.

\subsection{Stability of the nontransiting planet Kepler-160d}

We confirm the stability of the planetary system Kepler-160\,b/c upon inclusion of the nontransiting planet Kepler-160\,d near the period commensurabilites to planet~c modeled in Sect.~\ref{sec:non}. We integrated the system with the $N$-body code {\tt Mercury} \citep{1999MNRAS.304..793C} over 1\,Myr with a time step of 0.2~days (roughly a twentieth of the period of Kepler-160\,b). The system is more than a thousand times older than our integration time, but we had to limit ourselves to 1\,Myr to achieve results for the numerous configuration within a reasonable computation time. A more detailed stability analysis with longer integration times and over a range of orbital inclinations is reasonable when the number of possible orbital solutions has been narrowed down to a few or when a unique solution for the mass and period of Kepler-160\,d has been found.% All assumed system configurations remained stable for the 1~Myr.

\section{Conclusions}
\label{sec:conclusions}

We have analyzed the Kepler light curve of the solar-type star Kepler-160, which has been known to host two transiting planets. We detect a new sequence of three periodic transit-like features that we  attributed to a $1.9\,R_\oplus$ planet with virtually Earth-like insolation from a Sun-like star. The S/N of this signal is 9.5, the signal detection efficiency (SDE$_{\rm TLS}$) value is 16.3, the astrophysical FPP when the planet multiplicity boost is accounted for (FPP$_3$) is $1.81~{\times}~10^{-3}${, the multiple event statistic (MES) is 10.7, and the reliability against Kepler's rolling-band artifact and other instrumental artifacts is ${\sim}85\,\%$. We also showed that the signal is not caused by any nearby stellar contaminant.} All of these metrics point toward a genuine planet as the cause of the observed transit sequence. {For now, we remain cautious because there is mild evidence of the rolling-band artifact in the light curve and refer to it as a planet candidate KOI-456.04.} Our MCMC fits to the light curve suggest that this planet{, if real,} could have a global average surface temperature of around $+5^\circ$C assuming an Earth-like Bond albedo and an Earth-like atmospheric greenhouse effect. We present a list of {the next nine transit times until December 2028} (Table~\ref{tab:transits}). The next transit is projected to occur on 14.09.2020 (UTC).

We analyzed the previously known TTVs of Kepler-160\,c in trying to identify our new transit candidate as the source of the observed TTVs and thereby confirm it. Our MCMC analysis of this system with a three-body (one star plus two planets) TTV model did not result in a solution that involves the transiting candidate as a source of the observed TTVs of planet c, however. Instead, we found that Kepler-160 is orbited by a nontransiting planet, which we refer to as Kepler-160\,d. The sinusoidal nature of the observed TTVs suggests that Kepler-160\,d is close to a low-integer MMR with Kepler-160\,c. Under the assumption of a low mutual orbital inclination, we find that the resulting mass of this new planet Kepler-160\,d is always within the planetary mass range, with specific values between about 1 and 100 Earth masses.

\begin{acknowledgements}
{The authors thank Jeffrey Coughlin for helpful discussions of the planetary nature of the new transit signal. We also thank an anonymous referee for their valuable review of our manuscript.}
This research has made use of the NASA Exoplanet Archive, which is operated by the California Institute of Technology, under contract with the National Aeronautics and Space Administration under the Exoplanet Exploration Program. This work made use of NASA's ADS Bibliographic Services. This research has made use of ``Aladin sky atlas'' developed at CDS, Strasbourg Observatory, France. The Pan-STARRS1 Surveys (PS1) and the PS1 public science archive have been made possible through contributions by the Institute for Astronomy, the University of Hawaii, the Pan-STARRS Project Office, the Max-Planck Society and its participating institutes, the Max Planck Institute for Astronomy, Heidelberg and the Max Planck Institute for Extraterrestrial Physics, Garching, The Johns Hopkins University, Durham University, the University of Edinburgh, the Queen's University Belfast, the Harvard-Smithsonian Center for Astrophysics, the Las Cumbres Observatory Global Telescope Network Incorporated, the National Central University of Taiwan, the Space Telescope Science Institute, the National Aeronautics and Space Administration under Grant No. NNX08AR22G issued through the Planetary Science Division of the NASA Science Mission Directorate, the National Science Foundation Grant No. AST-1238877, the University of Maryland, Eotvos Lorand University (ELTE), the Los Alamos National Laboratory, and the Gordon and Betty Moore Foundation. This work has made use of data from the European Space Agency (ESA) mission
{\it Gaia} (\href{https://www.cosmos.esa.int/gaia}{www.cosmos.esa.int/gaia}), processed by the {\it Gaia} Data Processing and Analysis Consortium (DPAC, \href{https://www.cosmos.esa.int/web/gaia/dpac/consortium}{www.cosmos.esa.int/web/gaia/dpac/consortium}). Funding for the DPAC has been provided by national institutions, in particular the institutions participating in the {\it Gaia} Multilateral Agreement. RH is supported by the German space agency (Deutsches Zentrum f\"ur Luft- und Raumfahrt) under PLATO Data Center grant 50OO1501.
\end{acknowledgements}

\bibliographystyle{aa}
\bibliography{aa}

\begin{appendix}

\section{Detrending}
\label{sec:app_detrending}

%**********************************************
%Fig. A.1.
\begin{figure*}[h!]
\centering
\includegraphics[angle= 0,width=1\linewidth]{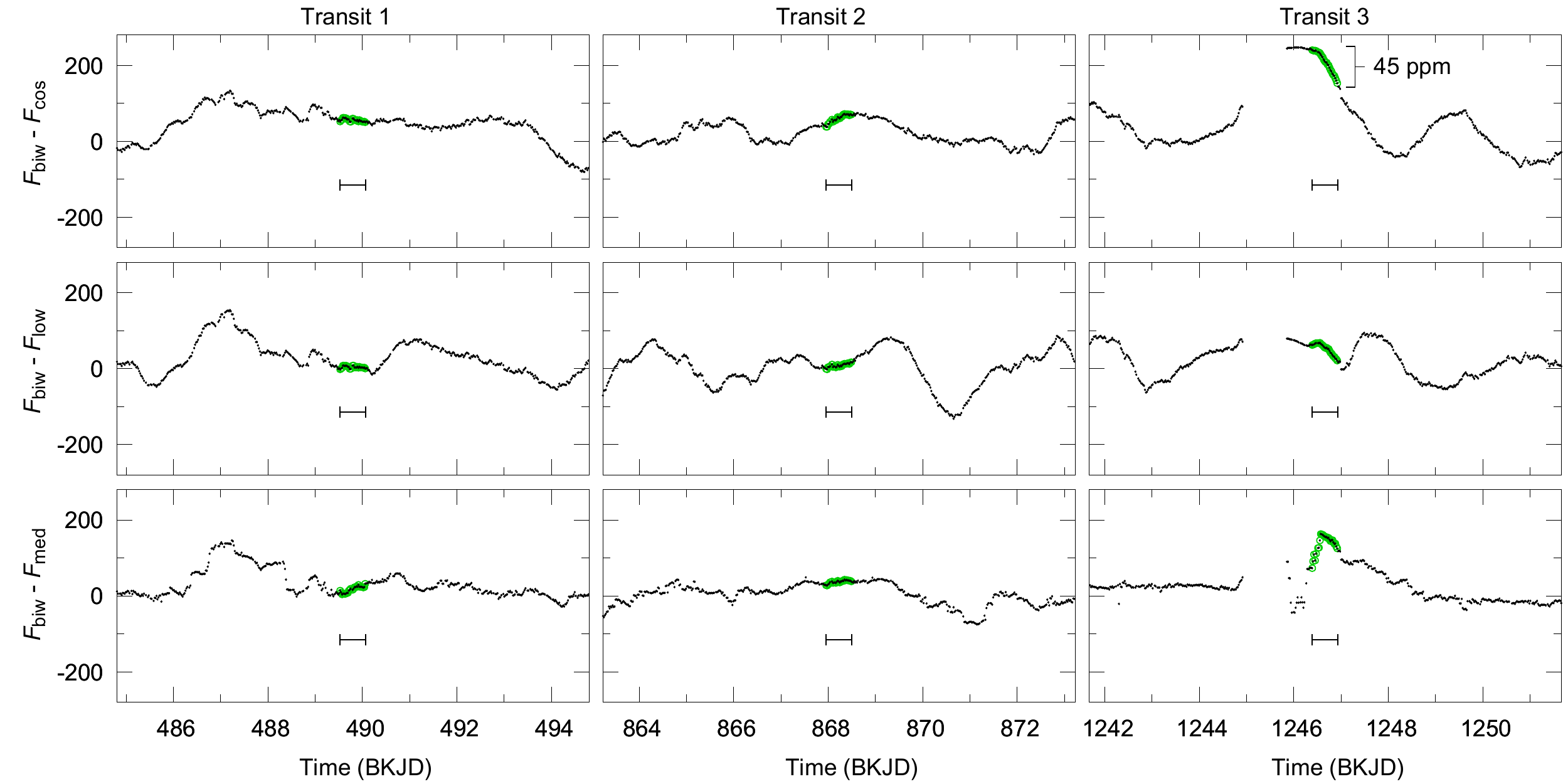}
\caption{Differential flux of different detrending methods compared to Tukey's biweight filter. The three panel columns refer to the three transits, and the in-transit data points, as detected by \texttt{TLS}, are highlighted in green as in Fig.~\ref{fig:lightcurve}(a)-(d). The detected transit width is also indicated with a horizontal bar in the center of each panel. The maximum in-transit differential flux that we found is 45\,ppm, which occurs during the third transit and for $F_{\rm biw}-F_{\rm cos}$ in the top right panel (see label). For comparison, the transit depth of our new candidate signal is about 300\,ppm.}
\label{fig:detrendings}
\end{figure*}
%**********************************************

Here we address the question of whether the new candidate signal that we detected might be due to the use of a particular detrending function. Prior to our nominal transit search and subsequent MCMC fitting of the light curve, we removed any nontransit variability on timescales longer than the expected transit duration (see Sect.~\ref{sec:detrending}) using the \texttt{W\={o}tan} implementation \citep{Hippke_2019} of Tukey's biweight filter \citep{mosteller1977data}. We refer to the resulting detrended flux as $F_{\rm biw}$, as shown in Fig.~\ref{fig:lightcurve}(e)-(g).

To explore the effect of the detrending algorithm on the transit signal, we also detrended the quarterly normalized PDCSAP flux (see Fig.~\ref{fig:lightcurve}a) using the other detrending algorithms from \texttt{W\={o}tan}. As examples, we focus on the cosine filter, resulting in a flux series $F_{\rm cos}$, the locally weighted or estimated scatter plot smoothing \citep[LOWESS;][]{Cleveland1979,Cleveland1981}, providing a flux series $F_{\rm low}$, and the running median filter as used in \cite{2019A&A...625A..31H}, producing a flux series $F_{\rm med}$. In Fig.~\ref{fig:detrendings} we show the differential flux series between the detrended light curve obtained with the biweight filter and  $F_{\rm cos}$ (top row), $F_{\rm low}$ (center row), and $F_{\rm med}$ (bottom row).

The overall variability of the differential flux is up to 200\,ppm, but mostly $<~100$\,ppm, on a timescale of about 10\,d. On timescales comparable to the duration of our new transit candidate signal of about 0.5\,d, however, the variability is typically on the order of a few times 10\,ppm. In particular, we find that the differential effect between the detrending methods is about 20\,ppm during the first and second transit both, and it is up to 45\,ppm during the third transit. This excursion is labeled in the top right panel in Fig.~\ref{fig:detrendings}. All things combined, the overall effect is small, although potentially not negligible, compared to the transit depth of about 300ppm.

Moreover, we find that the biweight filter that we used to prepare the light curve for our subsequent MCMC characterization of the system always resulted in transit depths that are a few times 10\,ppm shallower than obtained with the other detrending algorithms. As an example, we note that the differential in-transit flux during the third transit shown in Fig.~\ref{fig:detrendings} is slightly increased ($F_{\rm biw}-F_{\rm x}>0$), with $F_{\rm x}$ being any of the three flux series from the alternative detrending methods, which means that $F_{\rm biw}>F_{\rm x}$. In words, the other detrending algorithms produced slightly (${\sim}10$\,ppm) deeper signals during the third transit than the biweight filter that we used for our nominal system characterization. In other words, the transit depth (and likely also the signal detection efficiency) that we derived can be regarded as conservative estimates.

We conclude that the signal appears repeatedly even when the detrending filter is changed. The effect of the filter is about a few times 10\,ppm, which is small compared to the transit depth of about 300\,ppm.

\end{appendix}

\end{document}